\documentclass{fldauth}

\usepackage{mytex}
\usepackage{eepic}

\begin{document}

\FLD{1}{34}{0}{0}{0}

\runningheads{E. Kazantsev}
{Optimal Boundary Discretization by Variational Data Assimilation.}


\title{Optimal Boundary Discretization by Variational Data Assimilation.}

\author{ Eugene Kazantsev}

\address{
 INRIA, projet MOISE, 
 Laboratoire Jean Kuntzmann,\\
BP 53,
38041 Grenoble Cedex 9, 
France  }

\noreceived{}
\norevised{}
\noaccepted{}

\begin{abstract}
Variational data assimilation technique applied to the  identification of the optimal discretization of interpolation operators and  derivatives  in the nodes adjacent to the boundary of the domain  is discussed  in frames of the linear shallow water model. The advantage of controlling the discretization of operators near boundary rather than boundary conditions is shown.  
 Assimilating data produced by  the same model on a finer grid in a model on a coarse grid, we have shown that optimal discretization allows us to correct such errors of the  numerical scheme   as under-resolved boundary layer and wrong wave velocity.  

\end{abstract}




\section{Introduction}

It is now well known, that even the best model is not sufficient to make a good forecast. Any model depends on a number of parameters, requires initial and boundary conditions and other data that must be collected and used in the model.  However, interpolating or smoothing  observed data is not the best way to incorporate these data into a model. Lorenz, in his  pioneer work  \cite{Lor63} has shown that a geophysical fluid is extremely sensitive to initial conditions. This fact requires to bring  the model and its initial data together, in order to work with the couple "model-data", and to identify the optimal initial data for the model taking into account simultaneously the information contained in the observational data and in the equations of the model.

Optimal control methods \cite{Lions68} and perturbations theory   \cite{Marchuk75} applied to the data assimilation technique (\cite{Ledimet82}, \cite{ldt86}) show the way to do it. They allow to retrieve an optimal initial state for a given model from heterogeneous observation fields. Since the early 1990's, many mathematical and geophysical teams have been involved in the development of the data assimilation strategy. One can cite many papers devoted to this problem, as  in the domain of development of different techniques for the  data assimilation  and in the domain of its applications to the atmosphere and oceans.  

However,  overwhelming majority of data assimilation methods are now intended to identify and reconstruct an optimal initial state for the model. Since Lorenz \cite{Lor63}, who has pointed out  the importance of precise knowledge of the starting point of the model, essentially  the  starting point is considered as the control parameter and the target of  data assimilation. 
 
Of course, the model's flow  is extremely sensitive to its initial point. But, it is reasonable to suppose that a geophysical  model is also sensitive to many other parameters like bottom topography, boundary conditions on rigid and open boundaries, forcing fields and friction coefficients. All  these parameters and values  are also extracted in some way  from observational data, interpolated to the model's grid and can neither be considered as exact, nor as optimal to the model. On the other hand, due to non-linearity and intrinsic instability of model's trajectory, its sensitivity to all these external parameters may also be exponential. 

Numerous studies show strong dependence of the model's flow on the boundary data (\cite{VerronBlayo}, \cite{Adcroft98}), on the representation of the bottom topography (\cite{Holland73}, \cite{EbyHolloway}, \cite{LoschHeimbach}), on the wind stress (\cite{Bryan95}, \cite{Milliff98}), on diffusivity coefficients (\cite{Bryan87}) and on fundamental parametrization like Boussinesq and hydrostatic hypotheses \cite{LoschAdcroft}.  However, few papers are devoted to the development of data assimilation techniques intended  to identify and to control  these internal  model's parameters. One can cite several attempts to use data assimilation in order to identify the bottom topography of simple models (\cite{LoschWunsch}, \cite{assimtopo}) and in order to control open boundary conditions in coastal and regional models (\cite{shulman97}, \cite{shulman98}, \cite{Taillandier}). Boundary conditions on rigid boundaries have been controlled by data assimilation for heat equation (see for example \cite{ChenLin}, \cite{GillijnsDeMoor}), but this control concerns the linear parabolic  diffusion operator rather than hyperbolic transport and advection  operators that are more important in geophysical models.

 Studies of the possibility to control boundary conditions on rigid boundaries for equations containing hyperbolic operators can be found in  \cite{fxld-mo} on the example of non-linear balance equation, in  \cite{assimbc1} on the example of the wave equation and in  \cite{Leredde} and \cite{Lellouche} on the example of the Burgers equation.   The principal possibility to improve the model's solution controlling it's boundary values are shown in all these papers. However, as it has been noted in  \cite{Leredde}, particular attention must be paid to the discretization process which must respect several rules because it is the discretization of the model's operators takes into account the set of boundary conditions and introduces them into the model. Consequently, instead of controlling boundary conditions themself, it may be more useful to  identify optimal discretization of differential operators in points adjacent to  boundaries.  This allows  us to control directly the way the boundary conditions influence the model and to control boundary parameters in a more general way.   In  \cite{assimbc1}, for example, it was shown that deplacing  the boundary helps to correct the error occured due to a wrong wave velocity. Adjustment of the boundary position is, however, only possible when the coefficients used in approximations of the derivatives   are controlled directly by data assimilation.  Boundary conditions participate  in discretized operators, but considering the discretization itself, we take into account additional parameters like the position of the boundary,  lack of resolution of the grid, etc.  

Although the boundary configuration of the ocean is steady and can be measured with much better accuracy than the model's initial state, it is not obvious how to represent it  on the model's grid because of  limited resolution. The coastal line of continents possesses a very fine structure and can only be roughly approximated by the model's grid. Consequently,  boundary conditions  are defined at  the model grid's points which are different from the coast. Even the most evident impermeability condition being placed at a wrong point may lead to some error in the model's solution. We should accept the flux to be able to cross the boundary in places where the boundary is in water, prescribing some integral properties on the flux.

Moreover, ocean models frequently include strong and thin  boundary currents with intense velocity gradients.  In this case particular attention must be payed to the approximation of the   boundary layer because  wrong representation of these currents may be responsible for  drastic deformations in a global solution (see, for example \cite{VerronBlayo}). This may lead us to control the discretization of the model's operators in the whole boundary layer rather than  in  adjacent to boundary nodes only.

In this paper we use 4D-Var  data assimilation to control the discretization of derivatives and interpolation operators in the boundary regions. The development of the data assimilation is illustrated on the example of the linear shallow water model on the Arakawa's C-grid. The simplicity of the model allows us to clearly see technical points of the development (like the algorithm of differentiation and  development of the adjoint equation) without being overwhelmed by complexity of  operators and grids.  The purpose of the paper is to study the possibility to control the numerical scheme by data assimilation  and the particularities of this type of control in view to develop and use the data assimilation to identify optimal numerical scheme in coastal regions of ocean models.

The paper is organized as follows. The second section describes the model, its adjoint and the data assimilation procedure. The third section is devoted to the stationary solution of the model that includes the Munk boundary layer. And the fourth section discusses the data assimilation in the case of inertia-gravity and Rossby waves.

\section{Linear Shallow Water Model}

\subsection{Model's equations and discretization}
\label{sec1}

In order to test the data assimilation procedure we consider the linear shallow water model on the $\beta$-plane \cite{Gill}, \cite{Pedlosky}:
\beqr
\der{u}{t} - (f_0+\beta y) v&=& -g \der{\eta}{x} -\sigma u +\mu\Delta u +\fr{\tau_x}{\rho_0 H_0} 
\nonumber \\
\der{v}{t} + (f_0+\beta y) u&=& -g \der{\eta}{y} -\sigma v +\mu\Delta v +\fr{\tau_y}{\rho_0 H_0} 
\label{sw}  \\
\der{\eta}{t} &+& H_0(\der{u}{x}+\der{v}{y})=0 \nonumber
\eeqr
where $u(x,y,t)$ and $v(x,y,t)$ are two velocity components, $\eta(x,y,t)$ is the sea surface elevation, $\rho_0$ is the mean density of water, $H_0$ is the characteristic depth of the basin and  $g$ is the reduced gravity. The model is driven by the surface wind stress with components $\tau_x(x,y)$ and $\tau_y(x,y)$ and subjected to the  bottom drag that is parametrized by  linear terms $\sigma u,\; \sigma v$ and to the horizontal eddy diffusion parametrized by an  harmonic operator  $\mu\Delta u$ and $\mu\Delta v$. Coriolis parameter is supposed to be linear in $y$ coordinate and is presented as $(f_0+\beta y)$.  The system is defined in a square box of side $L$ with boundary conditions prescribed for $u$ and $v$. We require that   both  $u$ and $v$ vanish on the whole boundary. No boundary conditions is prescribed to $\eta$. 

We discretize all variables of this equation on the regular  Arakawa's C-grid with constant grid step $h=\fr{L}{N}$ in both $x$ and $y$ directions (see \rfg{grid})
\beqr
u_{i,j-1/2}&=&u(i h,j h-h/2) \mbox{ for } i=0,\ldots N, j=0,\ldots,N+1 \nonumber \\
v_{i-1/2,j}&=&v(ih-h/2,jh) \mbox{ for } i=0,\ldots N+1, j=0,\ldots N \nonumber \\
\eta_{i-1/2,j-1/2}&=&\eta(ih-h/2,jh-h/2) \mbox{ for } i=0,\ldots N+1, j=0,\ldots N+1 \nonumber 
\eeqr

\begin{figure}
\setlength{\unitlength}{1mm}
\newcount\indi
\newcount\indj
\newcount\num
\begin{center}
\begin{picture}(100,120)
\scriptsize
\multiput(10,10)(0,20){5}{\line(1,0){100}}
\multiput(10,10)(20,0){5}{\line(0,1){100}}
\Thicklines
\put(20,20){\line(1,0){90}}
\put(20,20){\line(0,1){90}}
\thicklines
\multiput(19,9)(20,0){5}{\line(1,1){2}}
\multiput(21,9)(20,0){5}{\line(-1,1){2}}
\indi=-1
\multiput(15,7)(20,0){5}{ 
\global\advance\indi by 1 $u$ }

\multiput(19,29)(20,0){5}{\line(1,1){2}}
\multiput(21,29)(20,0){5}{\line(-1,1){2}}
\indi=-1
\multiput(15,27)(20,0){5}{ 
\global\advance\indi by 1 $u$ }

\multiput(19,49)(20,0){5}{\line(1,1){2}}
\multiput(21,49)(20,0){5}{\line(-1,1){2}}
\indi=-1
\multiput(15,47)(20,0){5}{ 
\global\advance\indi by 1 $u$ }

\multiput(19,69)(20,0){5}{\line(1,1){2}}
\multiput(21,69)(20,0){5}{\line(-1,1){2}}
\indi=-1
\multiput(15,67)(20,0){5}{ 
\global\advance\indi by 1 $u$ }

\multiput(19,89)(20,0){5}{\line(1,1){2}}
\multiput(21,89)(20,0){5}{\line(-1,1){2}}
\indi=-1
\multiput(15,87)(20,0){5}{ 
\global\advance\indi by 1 $u$ }
\multiput(9,20)(0,20){5}{\line(1,0){2}}
\multiput(10,19)(0,20){5}{\line(0,1){2}}
\indj=-1
\multiput(10,17)(0,20){5}{ 
\global\advance\indj by 1 $v$ }

\multiput(29,20)(0,20){5}{\line(1,0){2}}
\multiput(30,19)(0,20){5}{\line(0,1){2}}
\indj=-1
\multiput(30,17)(0,20){5}{ 
\global\advance\indj by 1 $v$ }

\multiput(49,20)(0,20){5}{\line(1,0){2}}
\multiput(50,19)(0,20){5}{\line(0,1){2}}
\indj=-1
\multiput(50,17)(0,20){5}{ 
\global\advance\indj by 1 $v$ }

\multiput(69,20)(0,20){5}{\line(1,0){2}}
\multiput(70,19)(0,20){5}{\line(0,1){2}}
\indj=-1
\multiput(70,17)(0,20){5}{ 
\global\advance\indj by 1 $v$ }

\multiput(89,20)(0,20){5}{\line(1,0){2}}
\multiput(90,19)(0,20){5}{\line(0,1){2}}
\indj=-1
\multiput(90,17)(0,20){5}{ 
\global\advance\indj by 1 $v$ }

\multiput(10,10)(0,20){5}{\circle{2}}
\indj=-1
\multiput(6,11)(0,20){5}{ 
\global\advance\indj by 1 \num=\indj \multiply\num by 2 \global\advance\num by -1 $\eta$ }

\multiput(30,10)(0,20){5}{\circle{2}}
\indj=-1
\multiput(27,12)(0,20){5}{ 
\global\advance\indj by 1 \num=\indj \multiply\num by 2 \global\advance\num by -1 $\eta$ }

\multiput(50,10)(0,20){5}{\circle{2}}
\indj=-1
\multiput(47,12)(0,20){5}{ 
\global\advance\indj by 1 \num=\indj \multiply\num by 2 \global\advance\num by -1 $\eta$ }

\multiput(70,10)(0,20){5}{\circle{2}}
\indj=-1
\multiput(67,12)(0,20){5}{ 
\global\advance\indj by 1 \num=\indj \multiply\num by 2 \global\advance\num by -1 $\eta$ }

\multiput(90,10)(0,20){5}{\circle{2}}
\indj=-1
\multiput(87,12)(0,20){5}{ 
\global\advance\indj by 1 \num=\indj \multiply\num by 2 \global\advance\num by -1 $\eta$ }

\indj=-1
\multiput(7,2)(20,0){5}{ 
\global\advance\indj by 1 \num=\indj \multiply\num by 2 \global\advance\num by -1 $\the\num/2$ }
\indi=-1
\multiput(18,2)(20,0){5}{ 
\global\advance\indi by 1 $\the\indi$ }

\indj=-1
\multiput(0,9)(0,20){5}{ 
\global\advance\indj by 1 \num=\indj \multiply\num by 2 \global\advance\num by -1 $\the\num/2$ }

\indi=-1
\multiput(1,19)(0,20){5}{ 
\global\advance\indi by 1 $\the\indi$ }

\end{picture} 
\end{center}
\refstepcounter{fig}
\label{grid}
\caption{Arakawa C-grid}
\end{figure}
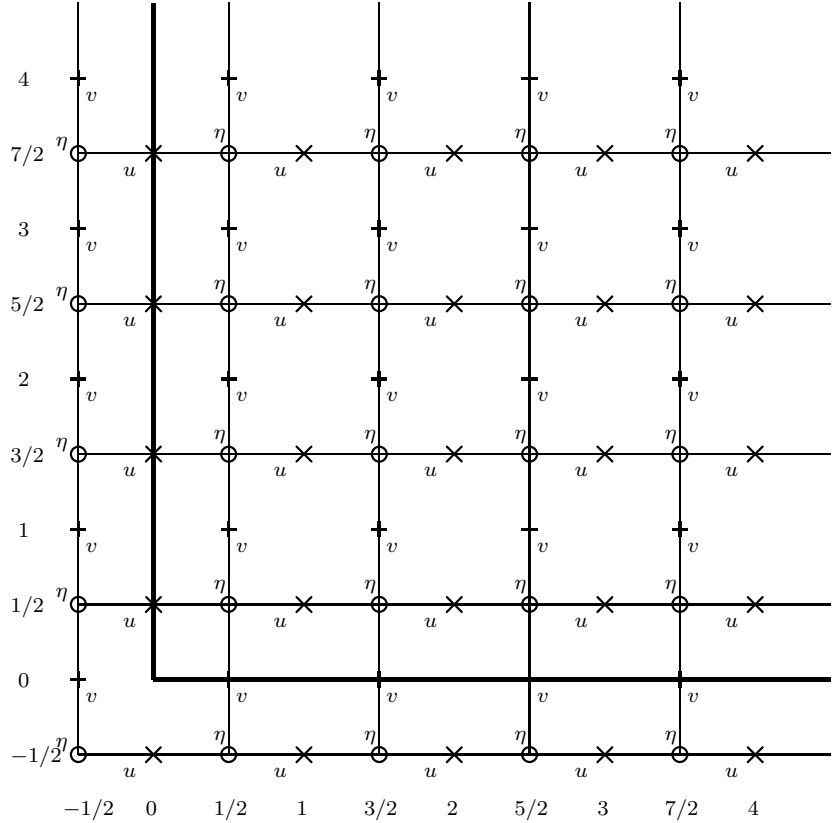

The system \rf{sw} contains several spatial operators  that must be approximated on this grid. First of all, this is the divergence operator in the third equation. We shall note the discretization of derivatives in the divergence by $D_x u$ and $ D_y v$. Second, the gradient operator that is present in the first two equations is denoted by $G_x \eta, G_y \eta $. Third, two interpolation operators  are necessary to calculate  variables $u$ and $v$ at points where $v$ and $u$ are defined respectively.  These operators are composed of two subsequent interpolations: $S_x u $ and $S_y v$ that provide their results at points where $\eta$ is defined, and $S_y \eta$ and $S_x \eta $ that give the resulting interpolation at  required points. It should be noted that operators $S_y \eta$ and $S_x \eta $ are never applied to the variable $\eta$ itself, but to the results of interpolation $S_x u $ and $S_y v$ which are defined at the same points as $\eta$. And finally,  the fourth operator is the discrete Laplacian that is denoted by $\Delta^h$.

Using these notations, the discretized system \rf{sw} can be  written
\beqr
\der{u}{t} &=& (f_0+\beta y) S_x S_y v-  g G_x\eta -\sigma u +\mu\Delta^h u +\fr{\tau_x}{\rho_0 H} 
\nonumber \\
\der{v}{t} &=&- (f_0+\beta y) S_y S_x u - g G_y \eta -\sigma v +\mu\Delta^h v +\fr{\tau_y}{\rho_0 H} 
 \label{sw-grid} \\
\der{\eta}{t} &=& - H_0(D_x u+D_y v) \nonumber
\eeqr

We define discretized operators $D_x u , D_y v, G_x \eta, G_y \eta, S_x u, S_y v, S_y \eta$ and $S_x \eta $ at all internal points of the domain as linear combination of the variable's values at four adjacent  grid points. For example,  the derivative and the interpolation of the variable $u$ defined at corresponding  points  writes
\beqr
(D_x u)_{i-1/2,j-1/2}&=&\fr{1}{h}\sum\limits_{k=-2}^{1} a^D_k u_{i+k,j-1/2}  \mbox{ for } i=M+1,\ldots , N-M
 \nonumber \\
 \label{intrnlsch} \\
(S_x u)_{i-1/2,j-1/2}&=&\fr{1}{h}\sum\limits_{k=-2}^{1} a^S_k u_{i+k,j-1/2}  \mbox{ for } i=M+1,\ldots , N-M \nonumber
\eeqr
That means the derivative at the point $i-1/2$ is presented as linear combination  of four values of $u$  taken at points from $i-2$ to $i+1$ with coefficients $a_k^D$.

 Coefficients $a_k$ are supposed to be known because we  intend to control the approximations near the boundary only. In this paper, we use either the sequence $ a^D_k=\fr{1}{h}(0,-1,1,0)$ for $ k=-2,-1,0,1$, or  $ a^D_k=\fr{1}{24h}(1,-27,27,-1)$. One can easily see that corresponding approximations of derivatives are of  the second and of the fourth order

\beqr 
\fr{u_{i}-u_{i-1}}{h}&=&\biggl(\der{u}{x}\biggr)_{i-1/2} + \fr{h^2}{24} \biggl(\tder{u}{x}\biggr)_{i-1/2}+ O(h^3) \nonumber\\
\label{taylor} \\
\fr{u_{i-2}-27u_{i-1}+27u_{i}-u_{i+1}}{24h}&=&\biggl(\der{u}{x}\biggr)_{i-1/2} - \fr{3 h^4}{640} \biggl(\pder{u}{x}\biggr)_{i-1/2} 
+ O(h^5) \nonumber\eeqr

To discretize  interpolation operators, we use either $ a^S_k=(0,1/2,1/2,0)$ for $ k=(-2,-1,0,1)$, or  $ a^S_k=\fr{1}{8}(-1,9,9,-1)$ that also provide the second and the fourth order interpolation of the variable. 
   
All other operators are discretized similarly at all internal points. All derivatives  use the same sequence $ a^D_k$ and all interpolation operators are discretized with the same sequence $ a^S_k$.  We shall note  in each particular case whether the second or the fourth order scheme is used. 

However, operators $D$ and $S$ are not  defined in the boundary region that is considered as a band of $M$ grid nodes width that follows the boundary.   Discretizations of operators in this band are supposed to be different from \rf{intrnlsch} and represent the control variables in this study. Moreover, schemes \rf{intrnlsch} can not be used at all for the fourth order approximation because they require function's values  beyond the boundary. We can, of course, extrapolate variables beyond the domain with the necessary order and substitute  extrapolated values in \rf{intrnlsch}, but it is not obvious what extrapolation formula is the best for this purpose. So, in order to obtain an optimal boundary approximation assimilating external data, we suppose nothing about derivatives near the boundary  and  write them in a general form
\beq
(D_x u)_{m-1/2,j-1/2}=\fr{1}{h}\biggl(\alpha_{0,m}^{D_xu}+\sum\limits_{k=1}^{K} \alpha^{D_xu}_{k,m} u_{k-1,j-1/2}\biggr) \mbox{ for } m=1,2,\ldots, M \label{bndsch}
\eeq
This formula represents also a linear combination of values of $u$ at $K$ points adjacent to the boundary, but with the coefficients $\alpha\neq a$ which are considered as particular for each operator and for each variable. The constant $\alpha_0^{D_xu}$, which is  also particular  for each operator, may be added in some cases to simulate non uniform boundary conditions like $u(0,y)=\alpha_0^{D_xu}\neq 0$.  All coefficients $\alpha$  are considered as control parameters in this study. They are allowed to vary in the variational assimilation procedure intended  to identify  their optimal set. The number of points  $K$ used in linear combination is not fixed. We keep the possibility to vary $K$ and  to study  its  influence on the assimilation results. In addition to this, we  can control derivative and interpolation operators at $M$ points adjacent to the boundary. The vicinity of the boundary is not restricted to just one point, we use the parameter $M$ to define how many points near the boundary will be controlled by the algorithm.  

Moreover, we can require looking for particular  $\alpha$ for each point near the boundary. That means we may allow the dependence of $\alpha_0,\; \alpha_{k,m}$ on the latitude $j$ in the expression \rf{bndsch}. The question of using either uniform coefficients or particular $\alpha$  for each latitude will be discussed later in this paper. 

Here, we can emphasize the choice of controlling the numerical scheme in the boundary region rather than boundary conditions.  The general form of boundary conditions that may be prescribed for $u$ variable in this model is
\beq
  u(t)\mid_{boundary}-r_1 \der{u}{x}(t)\biggr|_{boundary}=r_2.
\label{simplebc}
\eeq
 We can not write more complex boundary conditions (with second derivatives, for example) because we obtain a system with no solution at all.
Consequently, we can control only two parameters, $r_1$ and $r_2$. It may be sufficient in particular cases, but, as we shall see further, is not sufficient in general. Controlling all coefficients  of  the numerical scheme \rf{bndsch}, we are free to choose as many $\alpha_{k,m}$ as we need defining appropriate values of  parameters $K$ and $M$.

We distinguish $\alpha$ for different variables and different operators allowing different controls of  derivatives   because of the different nature of these variables and different boundary conditions prescribed for them. It is obvious, for example, that the approximation of the derivative of $\eta$ in the gradient may differ from the approximation of the derivative of $u$ in the divergence. Although both operators represent a derivative,  boundary conditions for $u$ and $\eta$ are different, these derivatives are defined at different points, at different distance from the boundary. Consequently, it is reasonable to let them be controlled separately and  to assume that their optimal approximation may be different with distinct coefficients $\alpha^{D_xu}$ and $\alpha^{D_x\eta}$.  

 Interpolations and derivatives on the  opposite side of the square are also assumed to be different. They are  calculated with coefficients $\tilde\alpha$ which are also considered as unknown control parameters:  
$$
(D_x u)_{N-m+1/2,j-1/2}=\fr{1}{h}\biggl(\tilde\alpha^{D_xu}_{0,m}+\sum\limits_{k=0}^{K} \tilde\alpha^{D_xu}_{k,m} u_{N-k+1/2,j-1/2}\biggr) \mbox{ for } m=1,2,\ldots, M
$$

Time stepping of this model is performed by the leap-frog scheme. 
The first time step is splitted into two Runge-Kutta stages in order to ensure the second order approximation.

\subsection{Tangent and adjoint models}

The approximation of the derivative introduced by \rf{intrnlsch} and \rf{bndsch} depends on control variables $\alpha$. Operators  are allowed to change their properties near boundaries in order to find the best fit with requirements of the model and data.   To assign  variables $\alpha$ we shall perform data assimilation procedure and find their optimal values.  Variational data assimilation is usually performed by minimization of the specially introduced cost function. The minimization is achieved using the gradient of the cost function that is usually determined by the run of the adjoint to the tangent linear model.

We start from development of the tangent linear model that is equal to the Gateaux derivative of the original model \rf{sw-grid} with respect to the control parameters. To calculate this derivative we suppose that the control variables can have small variations and we determine how these variations will perturb the model's solution. Thus, we suppose that all $\alpha$ are replaced by some $\alpha +\delta\alpha$ such that  $\norme{\delta\alpha} << \norme{\alpha}$. Let the model with $\alpha +\delta\alpha$  have a new solution $(u+\delta u, \; v+\delta v, \; \eta+\delta\eta)$. We get the equation for $\delta u, \; \delta v, \;\delta p$ as the difference between the model with coefficients $\alpha +\delta\alpha$  and the model with coefficients $\alpha $.   To obtain this equation, we introduce  variations of operators. Using the gradient $G$ as an example, we denote the difference
$ G(\alpha+\delta\alpha) - G(\alpha)$ by $\delta G$
and get
$$
 G(\alpha+\delta\alpha)(\eta+\delta\eta) - G(\alpha)\eta =(\delta G) \eta+ G \delta\eta + (\delta G) \delta\eta
$$ 
As it was defined, $G$ is a finite difference  approximation of a derivative. This matrix is block-bidiagonal (quadridiagonal if the fourth order scheme is used), i.e. each block is composed of two or fourth diagonals filled by coefficients $a^D_k$ from  \rf{intrnlsch}. However,  $M$ lines in the beginning and $M$ lines at the end  of each block have a different structure. They are intended to approximate  the gradient near the boundary and count each $K$ non-zero elements $\alpha$ as defined in \rf{bndsch}. It is these lines only that depend on the control parameter and  there are non-zero elements $\delta\alpha$ in the matrix $\delta G$ on these lines only. 
All other lines of $\delta G$ are filled by zeroes because the approximation of the derivative in the interior part of the domain does not depend on the controls $\alpha$. 
 
So far, both  $\delta G$  and $ \delta \eta$ are supposed to be small, we neglect their product in the right-hand-side of this equation.  
Variation of the divergence operator is written in a similar way with already neglected nonlinear term:
$$
 D_x(\alpha+\delta\alpha)(u+\delta u) - D_x(\alpha)u =(\delta D_x) u+ D_x \delta u 
$$ 
The interpolation operator  in \rf{sw-grid} is composed by two successive interpolations. The linear part of it's variation counts, consequently, three terms
$$
S_x(\alpha+\delta\alpha) S_y(\alpha+\delta\alpha) (v+\delta v) - S_x(\alpha) S_y(\alpha) v = (\delta S_x) S_y v+S_x (\delta S_y) v+S_x S_y \delta v
$$
All these matrices ($\delta S_x,\; \delta S_y,\; \delta D_x,\; \delta D_y $ etc.) have a similar structure. Their only non-zero elements are situated in the beginning and at the end of each block. 

Combining variances of approximations of  all differential and interpolation operators, we write the system that governs the evolution of perturbations of $u,v$ and $\eta$:

\beqr
\der{\delta u}{t} &=& \underline{(f_0+\beta y) S_x S_y \delta v-  g G_x\delta\eta  -\sigma \delta u +\mu\Delta^h \delta u} +
\nonumber \\
 &+&\underline{\underline{(f_0+\beta y)(\delta S_x S_y v+S_x \delta S_y v) -g\delta G_x\eta }}
\nonumber \\
\der{\delta v}{t} &=&\underline{- (f_0+\beta y) S_y S_x \delta u - g G_y \delta \eta -\sigma \delta v +\mu\Delta^h \delta v }-
\label{tlm} \\
&-&\underline{\underline{(f_0+\beta y) (\delta S_y S_x u+S_y \delta S_x u) -g\delta G_y \eta}}
\nonumber \\
\der{\delta\eta}{t} &=& - \underline{H_0(D_x \delta u+D_y \delta v)}- \underline{\underline{H_0(\delta D_x u+\delta D_y v)}} \nonumber 
\eeqr
with the same zero boundary conditions  for $\delta u$ and $\delta v$  as for $u$ and $v$. No boundary conditions is prescribed for  $\delta\eta$ as well as for $\eta$. At initial time both  $\delta u, \delta v $ and $ \delta \eta$ vanish because our study is confined at the control of the  discretization of operators near the boundary only rather than joint control of a boundary scheme and initial conditions of the model. We are interested in the evolution of a pure perturbation due to boundary scheme.  

We have intentionally distinguished two types of operators in the right-hand-side of \rf{tlm} (underlined by one or two lines). Operators of the first type act in the space of perturbations of $u,v$ and $\eta$. Both the argument and the result of these operators are in the space of $\delta u,\; \delta v,\; \delta\eta$. Operator $G_x$, for example, brings the variable $\delta\eta$ to the space of $\delta u$, operator  $S_y S_x$ transforms the variable $\delta u$ to the space of $\delta v$, etc. 

The second type of operators distinguished in \rf{tlm} (underlined by two parallel lines) are of implicit structure. Their argument  in the present form is contained in the matrix itself. Let us consider, for example, the  expression $\delta G_x \eta $. True argument of this expression is contained in the matrix $\delta G_x$ rather than in the vector $\eta$. In this form, it is the matrix that depends on the variations of the control parameter $\delta\alpha$. The vector $\eta$,  that can be wrongly considered as an argument  in this expression,  is the solution of the original shallow-water model \rf{sw}. This vector depends implicitly on the control parameters $\alpha$, but does not depend on variations $\delta\alpha$. 

Expressions like $\delta G_x \eta $ are not convenient to carry out  further development. Writing an adjoint operator, we would better have a constant operator (which does not depend on $\delta\alpha$)   multiplied by a variable  vector (which depends on $\delta\alpha$). It would be more convenient to  rewrite  products like this in  a more explicit form
$$\delta G_x \eta=\widehat\eta\vec\delta\alpha$$ 
where the  operator $\hat \eta$ is constructed from the solution $\eta$  of the original equation  and the vector $\vec{\delta\alpha}$ is extracted from the matrix $\delta G$ as it is shown in \cite{assimbc1}. 
All other products of the second type ($ \delta D_x u\; \delta D_y v,\; \delta S_x u,$ etc) are also rewritten in this, more explicit form. We shall further use hats to denote matrices that have been constructed from vectors. These matrices are also block-matrices. All elements of their blocks are equal to zero except $M$ lines in the beginning and $M$ lines at the end  of each block. These lines are composed of $K$ values of corresponding vector, and, namely, values of approximated function in $K$ nodes  near the boundary.

In order to write the tangent model in a matrix form we add one additional equation to the system \rf{tlm} expressing that control coefficients $\alpha$ are stationary: $\der{\delta\alpha}{t}=0$. Using these notations we write:
\beq
\der{}{t}
\left(\begin{array}{c}
 \delta u\\ \delta v\\ \delta\eta\\ \delta\alpha
\end{array}\right)=
\left(\begin{array}{cccc}
-\sigma+\mu\Delta^h    &(f_0+\beta y) S_x S_y& -g G_x&R_u\\
- (f_0+\beta y) S_y S_x& -\sigma+\mu\Delta^h &-g G_y&R_v\\
- H_0 D_x&- H_0 D_y& 0&R_\eta\\
0&0&0&0
\end{array}\right)
\left(\begin{array}{c}
 \delta u\\ \delta v\\ \delta\eta\\ \delta\alpha
\end{array}\right) \label{tlm-mat}
\eeq
where
\beqr
R_u\delta\alpha&=& (f_0+\beta y)(\delta S_x S_y v+S_x \delta S_y v) -g\delta G_x\eta =[(f_0+\beta y)(\widehat{S_y v}+S_x  \widehat{v}) -g\widehat\eta]\delta\alpha \nonumber \\
R_v\delta\alpha&=&-(f_0+\beta y) (\delta S_y S_x u+S_y \delta S_x u) -g\delta G_y \eta=[-(f_0+\beta y) (\widehat{S_x u}+S_y \widehat{u}) -g\widehat{ \eta}]\delta\alpha 
\nonumber \\
R_\eta\delta\alpha&=&[- H_0(\delta D_x u+\delta D_y v)=- H_0(\widehat{u}+\widehat{v})]\delta\alpha\label{R}  
\eeqr

It has to be noted, that operators $R_u,\; R_v$ and $R_\eta$ act from the space of the control variable $\alpha$ to the space ofvariations of the  model's solution $\delta u,\; \delta v$ or $\delta \eta$. Their matrices, consequently,  are  rectangular. The number of lines in their matrices corresponds to  dimensions of variables in physical space, i.e. $N\tm (N-1)$ for $u$ and $v$ and $(N-1)^2$ for $\eta$. The number of columns corresponds to the dimension of the control variable's space or to the total number of control coefficients $\alpha$. This number is equal to the number of controlled operators (eight in the present configuration) multiplied by the number of boundaries to control for each operator (usually two: either East-West if the operator is in $x$ direction, or North-South if the operator acts in $y$) and multiplied by the number of control coefficients for each boundary ($M\tm (K+1)$ as defined by \rf{bndsch}).  The smallest number of control coefficients in the present configuration is equal to 48 ($K=2,\;M=1$). Of course, this small number is not always sufficient. In general case, we have to use  more controls to get better results, but in any case the number of control coefficients is much less than the dimension of the variables space.  

This fact may be taken into account in the analysis of the complexity of variational data assimilation in this case and in the making choice of the practical way of calculation of the cost function. 
We can see, the matrix of the  tangent linear model \rf{tlm-mat} is  composed by two  parts: the $3\tm3$ block composed of operators acting in the space of the model's variables and the fourth column composed of operators $R$.  The $3\tm3$ block is responsible for the evolution of a small perturbation by the model's dynamics, while the column  determines the way how this perturbation is introduced into the model. The first term is similar for any data assimilation, while the second one is specific to the particular variable under control. This term is absent when the goal is to identify the initial conditions of the model because the uncertainty in initial conditions  is introduced only once, at the beginning of the model integration. But, when the uncertainty is presented in an internal model parameter, like in this case, the perturbation is introduced at each time step of  the model. 

So, if we intend to identify an optimal boundary parametrization for a model with an existing adjoint developed for data assimilation and identification of its initial state, we can use this adjoint as $3\tm3$ block because this part is common for any data assimilation.  However, the fourth column of the matrix \rf{tlm-mat}  must be developed from the beginning because it is specific to the particular control parameter. 
This development may be technically difficult. One can see, writing adjoint of operators $R$ \rf{R} is  as complex as the adjoint of the $3\tm3$ block.   If we add nonlinear terms to the model, writing the fourth column will become even more complex than the development of the $3\tm3$ block. Including, for example, $v\;\der{u}{y} = S_x S_y v\; S_y D_y u$ term in the first model's equation will add six additional terms in the tangent model: two terms in the $3\tm3$ block and four  terms in the fourth column. 

Regarding the supplementary work to be done developing the tangent and adjoint models and taking into account small number of control parameters,  it may be reasonable to try to calculate the gradient of the cost function by some other method beginning with  the simplest finite difference method. Of course, this will be more expensive computationally, but the gain in the development procedure may compensate this excessive computational cost. 

Moreover, boundary conditions and discretizations of operators near boundary are internal model's parameters. They must be identified once for all model runs,  while initial conditions are external parameters and must be identified for each particular model run. Consequently, we have to identify the discretization near boundary on the stage of the development of the model while choosing all other internal parameters. Hence, we can accept more time consuming data assimilation procedure under condition that it  requires less time for it's development.

However, in this paper we proceed with the calculation of the gradient of the cost function using the adjoint of the tangent linear model \rf{tlm-mat}. In order to develop the adjoint model, we need to introduce the scalar product in the space defined by tangent model. Each element in this space is composed of discretized functions $ u,\; v$ and $\eta$ and also the whole set   of the control coefficients $\alpha$. A vector in this space has four components $\phi=( u, v, \eta, \alpha)$. 

In this paper we consider a weighted  Euclidien scalar product in this space 
\beqr
\spm{\phi}{\phi^*}&=&\spm{\left(\begin{array}{c}
   u\\   v\\  \eta\\  \alpha
\end{array}\right) }{
 \left(\begin{array}{c}
  u^*\\  v^*\\ \eta^*\\ \alpha^*
\end{array}\right)} = \label{sp}\\
&=& w_u^2\sum_{i,j} u_{i,j} u^*_{i,j} + w_v^2\sum_{i,j} v_{i,j} v^*_{i,j}+w_\eta^2\sum_{i,j} \eta_{i,j} \eta^*_{i,j} +\sum_{k,m,operator} \alpha_{k,m}^{operator} (\alpha^{operator}_{k,m})^*
\nonumber
\eeqr
The sums in the scalar product is performed over all nodes $i,j$  of the grid  of all model's variables $u,\;v$ and $\eta$. The sum of control coefficients $\alpha$ is performed over all $k,m: 1\leq k\leq K,\; 1\leq m\leq M$ \rf{bndsch}  and over all operators ``$operator$'' controlled by these coefficients.
Weights $w_u=w_v=\fr{1}{\sqrt{gH_0}}$ and $w_{\eta}=\fr{1}{H_0}$ are introduced to  bring all variables to dimensionless form.

Denoting the matrix of the tangent model \rf{tlm-mat} by $A$, we write it in a short form:
$$ \der{\delta\phi}{t}-A\delta\phi=0,\quad 
\delta\phi\mid_{t=0}=
\left(\begin{array}{c}
 0\\0\\0\\ \delta\alpha
\end{array}\right)
$$
Following \cite{Marchuk75},  \cite{ldt86}, we multiply the tangent model  by some $\phi^*(t)$ and integrate this product in time from 0 to some $T$. 
\beqr
0&=&\int\limits_0^T \spm{(\der{\delta\phi}{t}-A\delta\phi)} {\phi^*(t)} dt = 
\int\limits_0^T \spm{\der{\delta\phi}{t}} {\phi^*(t)}dt - \int\limits_0^T \spm{A\delta\phi}{ \phi^*} dt =
\nonumber\\
&=&\spm{\delta\phi(T)}{\phi^*(T)}-\spm{\delta\phi(0)}{\phi^*(0)}+ 
    \int\limits_0^T \spm{\delta\phi(t)}{-\der{\phi^*}{t}- A^*\phi^*} dt 
\eeqr
where  $A^*$ is the adjoint matrix to $A$.  
Consequently, if $\phi^*$ satisfies the adjoint model $-\der{\phi^*}{t}- A^*\phi^*=0$, then we have an equality 
\beq
\spm{\delta\phi(T)}{\phi^*(T)}=\spm{\delta\phi(0)}{\phi^*(0)} \label{equality}
\eeq
that we shall use to calculate the gradient of the cost function. 

Complete matricial formulation of the   adjoint model, hence,     writes:
\beq
-\der{}{t}
\left(\begin{array}{c}
  u^*\\  v^*\\ \eta^*\\ \alpha^*
\end{array}\right)=
\left(\begin{array}{cccc}
-\sigma+\mu\Delta^h    &(f_0+\beta y) S_x^* S_y^*& -H_0 D_x^*&0\\
- (f_0+\beta y) S_y^* S_x^*& -\sigma+\mu\Delta^h&-H_0 D_y^*&0\\
g G_x^*& g G_y^*& 0&0\\
R_u^*&R_v^*&R_\eta^*&0
\end{array}\right)
\left(\begin{array}{c}
  u^*\\  v^*\\ \eta^*\\ \alpha^*
\end{array}\right) \label{am}
\eeq
where
\beqr
R_u^*&=&  (f_0+\beta y)(\widehat{S_y v}^*+ \widehat{v}^*S_x^* ) +g\widehat\eta^* \nonumber \\
R_v^*&=&-(f_0+\beta y) (\widehat{S_x u}^*+\widehat{u}^*S_y^* ) +g\widehat{ \eta}^* \label{R*} \\
R_\eta^*&=&- H_0(\widehat{u}^*+\widehat{v}^*) \nonumber 
\eeqr

The result of the adjoint model run $\phi^*(0)$, that is used in \rf{equality}, we shall denote as
\beq\phi^*(0)={\cal A}^*(T,0)\phi^*(T)\label{am-mat}\eeq
in order to explicitly show that it was obtained by  the adjoint model \rf{am} that starts at time $t=T$ from the state $\phi^*(T)$ and is integrated back in time to $t=0$. 

\subsection{Cost function}

One of principal  purposes of variational data assimilation consists in the variation of  control parameters in order to bring the model's solution closer to the observational data. This implies the necessity to measure the distance between the trajectory of the model and data. Introducing a cost function, we define this measure. Generally speaking, the cost function is represented by some norm of the difference between model's solutions and observations, eventually accompanied by some regularization term. 

In this paper we use the  Euclidean norm  because in this preliminary study we are interested essentially in the principal possibility to control the boundaries by data assimilation, rather than to the difficulties that arise due to imperfect data. Consequently, all the artificial observations to be assimilated into the model are considered as perfect, known at any time and at any point for all variables $u,\; v$ and $\eta$,  
 and  no regularization is added to the cost function. To characterize the difference between the model's solution and  the observational data we use
\beq 
\xi^2=w_u^2\sum_{i,j} (u_{i,j}- u^{obs}_{i,j})^2 + w_v^2\sum_{i,j}(v_{i,j}- v^{obs}_{i,j})^2+w_\eta^2\sum_{i,j} (\eta_{i,j}- \eta^{obs}_{i,j})^2 \label{xi}
\eeq
The norm $\xi$ is, in fact,  associated with the scalar product \rf{sp} with the fourth component $\alpha$ equal to zero. Expressing $\xi$ in terms of the scalar product, we emphasize its dependence on time and control coefficients $\alpha$:
\beqr
\xi^2&=&\xi^2(\alpha,t)=\spm{\phi(\alpha,t)-\phi^{obs}(t)}{\phi(\alpha,t)-\phi^{obs}(t)}=
\nonumber\\
&=&\spm{\left(\begin{array}{c}
   u(\alpha,t)-u^{obs}(t)\\   v(\alpha,t)-v^{obs}(t)\\  \eta(\alpha,t)-\eta^{obs}(t)\\  0
\end{array}\right) }{\left(\begin{array}{c}
   u(\alpha,t)-u^{obs}(t)\\   v(\alpha,t)-v^{obs}(t)\\  \eta(\alpha,t)-\eta^{obs}(t)\\  0
\end{array}\right) 
 }
\eeqr

The choice of the cost function for variational data assimilation performed to identify discretizations of model's operators near boundaries is not obvious. If the purpose is to identify the initial conditions of the model, it is reasonable to choose a classical 4D-VAR cost function 
\beq
\costfun_4(\alpha)=\int\limits_0^T \xi^2(\alpha,t)  dt \label{costfn-4}
\eeq
that gives the same importance to the difference $\xi^2$ at any time. 
However, assimilating data with the purpose to identify an internal model's parameter, it may be more reasonable to emphasize the end of the assimilation window $[0,T]$ with respect to it's beginning because in the beginning of the model's integration the value of $\xi(t)$ is low and less important. Indeed, supposing observational data to be perfect, we start from the exact initial state obtained from artificial observations and  get $\xi(0)=0$.  

Consequently, we shall try to use two other cost functions 
\beq
\costfun_{fp}(\alpha)=\xi^2(\alpha,T),\;\;\; \costfun_{T}(\alpha)=\int\limits_0^T t \xi^2(\alpha,t)  dt \label{costfn-fpT}
\eeq
 requiring that the end of the interval is weighted more heavily. 

 To calculate the gradient of the cost function, we  calculate  first its variation:  
 \beqr
 \delta\costfun_{fp}&=&\costfun_{fp}(\alpha+\delta\alpha)-\costfun_{fp}(\alpha)=  \xi^2(\alpha+\delta\alpha,T)-\xi^2(\alpha,T)=
  \nonumber \\
&=&\spm{\phi(\alpha+\delta\alpha,T)-\phi^{obs}(T)}{\phi(\alpha+\delta\alpha,T)-\phi^{obs}(T)}-\nonumber \\
& &-\spm{\phi(\alpha,T)-\phi^{obs}(T)}{\phi(\alpha,T)-\phi^{obs}(T)} \sim
  \nonumber \\
&\sim& 2\spm{\delta\phi(T)}{\phi(\alpha,T)-\phi^{obs}(T)}
\eeqr 
This scalar product is similar to the scalar product  in the left-hand-side of \rf{equality} with   $\phi^*(T)$ replaced by $\phi(\alpha,T)-\phi^{obs}(T)$. Consequently, if we run the adjoint model \rf{am}  from the state $\phi^*(T) = \phi(\alpha,T)-\phi^{obs}(T)$ back in time from $t=T$ to $t=0$ we obtain the state $\phi^*(0)$. This state, being scalarly multiplied by $\delta\phi(0)$ provides the variation of the cost function
\beq
\delta\costfun_{fp}= 2\spm{\delta\phi(0)}{\phi^*(0)}
\eeq
As it has been noted,  first three components of $\delta\phi(0)$ are equal to zero. Consequently, first three components of $\phi^*(0)$  are multiplied by zero and we are interested only in the fourth component of the result of the adjoint model $\alpha^*$. 

 Thus, the gradient of the cost function is related  to the fourth component of  $\phi^*(0)$
 \beq
 \nabla \costfun_{fp} = 2  [\phi^*(0)]_4 =2[{\cal A}^*(T,0) (\phi(\alpha,T)-\phi^{obs}(T))]_4
 \label{grad-fp}
 \eeq
and namely to the  $\alpha^*(0)$ is obtained as the final point  of the adjoint model integrations from $t=T$ to $t=0$ beginning with the state $\phi^*(T) = \phi(\alpha,T)-\phi^{obs}$. 

Gradients of two other cost functions can be obtained in a  similar way
\beqr
\nabla \costfun_{4} &=&  2\int\limits_0^T {\cal{A}}^*(T,t) (\phi(\alpha,t)-\phi^{obs}(t))dt\nonumber\\
\nabla \costfun_{T} &=&  2\int\limits_0^T t {\cal{A}}^*(T,t) (\phi(\alpha,t)-\phi^{obs}(t))dt
 \label{grad}
 \eeqr

These gradients are used in the minimization procedure that is implemented in order  to find the minimum
\beq
\costfun(\bar\alpha) = \min_{\alpha} \costfun(\alpha)
\eeq
Coefficients $\bar\alpha$  are considered as coefficients realizing  optimal discretization of the model's operators in the boundary regions. 

The  minimization procedure used here was developed by Jean Charles Gilbert and  Claude Lemarechal, INRIA \cite{lemarechal}.  The procedure uses the limited memory quasi-Newton method.

\section{Stationary solution of the model}

In this section we perform several experiments with the stationary model's solution that contains a strong boundary layer. The model is considered in a square box of side length $L=4000$ km driven by a steady, zonal  wind forcing with now classical sinusoidal profile
$$
\tau_x=\tau_0 \cos \fr{2\pi (y-L/2)}{L}
$$
that leads to the formation of a double gyre circulation \cite{LPV}. The maximal wind tension on the surface is taken to be $\tau_0=1.5\fr{dyne}{cm^2}$. 

As it has been already noted, the Coriolis parameter is linear function in $y$ with  $f_0=10^{-4}s^{-1}$ and $\beta=2\tm 10^{-11} (ms)^{-1}$. The reduced gravity and the depth are respectively equal to $g=0.02\fr{m}{s^2},\;H_0=1000m$.  The coefficient of Eckman dissipation is chosen as
$\sigma=5\tm 10^{-8}s^{-1}$, that corresponds to the damping time-scale 
$ T_\sigma = 2\times10^7s \sim 200\mbox{ days}.$ The lateral friction coefficient $\mu$ has  been taken to be $\mu=800\fr{m^2}{s}$, that corresponds to the damping time scale $T_{\mu}= 2.5$ days for a  wave of 100 km length.  

All operators in the model are approximated with the second order accuracy both in the interior of the domain and near its boundary. That means all the coefficients $a$  have been chosen to provide second order discretization as it has been discussed in the section \ref{sec1}. The expression \rf{bndsch}, that is used to interpolate functions and to calculate their derivatives near boundary, is written with $K=2,\; M=1,\; \alpha_0=0$. Coefficients $\alpha$ are defined as $\alpha_{1,1}=-1/h,\; \alpha_{2,1}=1/h$ for all derivative operators and $\alpha_{1,1}=\alpha_{2,1}=1/2$ for all interpolations. That gives, for example, the value of the derivative of $u$ at the point $i=1/2$ as $(D_x u)_{1/2,j-1/2}=\fr{u_{1,j-1/2}-u_{0,j-1/2}}{h}   $.

The  stationary solution of the model in this configuration possesses a boundary layer near the Western boundary. This is a  well known Munk layer \cite{Munk}  that is characterized by the local equilibrium between the $\beta$-effect and the lateral dissipation. It's width can be easily calculated by solving the Munk equation

\beq
\tder{v}{x}-\fr{1}{\delta^3}v=0, \;\; v(0)=0,\;v(\infty)=0,\;\delta=\biggl(\fr{\mu}{\beta}\biggr)^{1/3}
\label{Munk}
\eeq
The solution has a form
$$v(x)=Ce^{\fr{-x}{2\delta}}\sin(\fr{\sqrt{3}x}{2\delta})$$

The width of the Munk layer for given parameters is equal to $\fr{2\pi}{\sqrt{3}}(\mu/\beta)^{1/3}=124$ km.

 As it has been  discussed in \cite{Bryan75}, the model must resolve this layer with at least one grid point (optimally more than one grid point) in order to maintain numerical stability.  The work of \cite{Griffies} emphasized the importance of having at least two grid points in the Munk layer in order to minimize the level of spurious oscillations visible in the velocity fields as well in the field of the sea surface elevation. 

The resolution of the model in this section is intentionally chosen to be too coarse to resolve the Munk layer. The model's grid is composed of 30 nodes in each direction, that means the grid-step is equal to 133 km, that is more than  the Munk layer's width. As it can be seen in \rfg{grid}, there is only one grid node in the layer for variables $v$ and $\eta$ and no nodes for $u$ variable. 

Artificial ``observational`` data are generated by
the same model with all the same parameters but with 9 times finer resolution  (15 km  grid step). The model with fine resolution, having 8 nodes in the Munk layer, resolves  explicitly the layer and must have no spurious oscillations. All nodes of the coarse grid belong to the fine grid, consequently, we do not need to interpolate ''observational'' data to the coarse grid. We just take values in  nodes of the high resolution grid that correspond to nodes on the coarse grid. 

Along with the models on the fine grid that is used to generate ``observations'' and on the coarse grid that is used to perform data assimilation experiments, we also use the medium resolution model on the grid 3 times finer than the coarse grid. The grid step of the medium grid is equal to  44 km, there are 3 nodes in  the Munk layer, which is, hence, relatively well resolved.  

In order to compare these three models, we use the $v$ velocity field because under-resolution of the Munk layer in the most  visible in the field of this variable. If we compare profiles of the velocity $v(x,y)$ plotted in \rfg{profil} at $y=1000$ km (one quarter of the side length), we see that the profiles of the  models with fine and medium resolutions coincide almost everywhere. The only visible difference is seen between the second and the third nodes of the medium-resolution grid. The maximum of velocity is located just between these nodes and is not reproduced on the medium grid. However, values of the velocity in nodes adjacent to the maximum are close to corresponding values of the model on the fine grid. 

The profile obtained with the coarse resolution model (solid line in \rfg{profil}) shows  strong spurious oscillations due to the unresolved Munk layer.  As expected, one point in the layer is not sufficient to suppress the numerical mode in the velocity field. Similar oscillations are also present in the fields of the velocity $u$ and SSH  $\eta$.

\begin{figure}
  \begin{center}
  \begin{minipage}[r]{1\textwidth} 
  \centerline{\includegraphics[angle=0,width=0.79\textwidth]{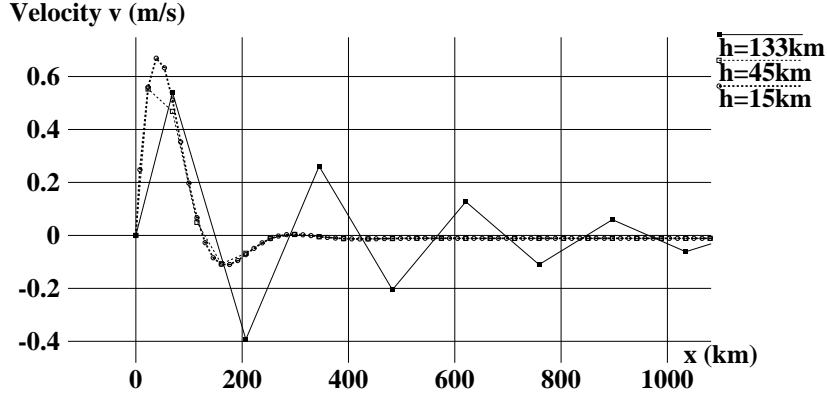}}
  \caption{ Profiles of the velocity $v(x,y)$ at $y=1000$ km obtained on the coarse grid with $h=133km$ (solid line), medium grid with $h=45$ km (dashed line) and fine grid with $h=15$ km (bold dashed line) }
  \end{minipage} 
  \end{center} 
\refstepcounter{fig}
\label{profil}
\end{figure}

Performing the data assimilation we hope that the  optimal discretization of operators near the boundary may help us to suppress these spurious oscillations even working on the coarse grid.

\subsection{Parameters of the assimilation procedure}

In this section we shall discuss such parameters of the assimilation procedure as the length of the assimilation window, the choice of the cost function among $\costfun_4,\; \costfun_{fp}$ and $\costfun_{T}$ and the physical meaning of the control parameters and their eventual dependence on the coordinate. 
As the first experiment, we identify only three coefficients for each operator and each boundary. That means, we take $K=2,\;M=1$ in \rf{bndsch}.

We can note that this choice is almost equivalent to the control of the boundary conditions prescribed for $u$ and $v$ variables.   Indeed, supposing, for example, the boundary conditions for $u(x,y)$ for theoperator $\der{u}{x}$   at $x=0$ are expressed according to \rf{simplebc}, the derivative of $u$  at $x=h/2$ that we need in the third equation of the model \rf{sw} can be found as $ \der{u}{x}\biggl|_{1/2}=\fr{u_{1,j}}{h-r_1} - \fr{r_2}{h-r_1}$.  Therefore, if we want to control boundary conditions, we can control either $r_1$ and $r_2$ in the expression of boundary conditions or $\alpha^{D_xu}_{0,1}=- \fr{r_2}{h-r_1}$ and $ \alpha^{D_xu}_{2,1}=\fr{1}{h-r_1}$ in discretization of the operator \rf{bndsch}. The result will be the same in both cases: we control two parameters looking for minimizing the same cost function (the coefficient $\alpha^{D_xu}_{1,1}$ is not controlled due to keeping formal Dirichlet condition  $u(0,y)=0$). 

However, even in this case there is a difference between controlling the boundary conditions and controlling the discretizations of operators. Controlling the boundary conditions and finding optimal $r_1$ and $r_2$ suppose using the same $r_1$ and $r_2$  for the interpolation operator near boundary. But finding the optimal $\alpha$, we suppose that $\alpha^{D_xu}$ and $\alpha^{S_xu}$ are independent on each other. 

Moreover, we admit that  coefficients in the gradient are also controlled by data assimilation,  despite no boundary condition is prescribed for  $\eta$.  This provide additional control that may help to improve the solution near the boundary.

From the physical point of view, assuming variable coefficients (either $r_1$ and $r_2$ or  $\alpha^{D_xu}_{0,1}$ and $ \alpha^{D_xu}_{2,1}$ we accept the flux can cross the boundary. Impermeability condition  $u(0,y)=0$ is no longer maintained. This fact must be addressed in the first turn, because it represents a major violation of the model's physics. Allowing the fluid to leave the domain in some place, we must allow it to come back in some other place in order to be able to keep the mass constant. In the present formulation, the coefficient $\alpha^{D_xu}_{0,1}$, that determines non null flux crossing the boundary,  is constant. That means, the flow may cross the boundary in the same direction all the boundary's long. The flux can not change  direction and reenter the domain in some place after leaving it in some other place.  Consequently, in order to allow the flux to come back,  we must at least allow  control coefficients  $\alpha_0$ to depend on the coordinates: latitude $y$ corresponding to the index $j$ for operators $S_x,\; D_x$ and $G_x$, and longitude $x$ or index $i$ for operators $S_y,\; D_y$ and $G_y$. Thus, we obtain a particular discretization of the operators  at each point.  Parameter $\alpha_{0,m}^{D_xu}$ in the expression \rf{bndsch} will have a form $(\alpha_{0,m}^{D_xu})_j$. We can also allow   all  parameters $\alpha_{k,m}$ to  depend on the coordinates. 

The initial guess for the minimization procedure is taken as the classical second order approximation, the same as the approximation that has been used to run the fine resolution model to generate artificial ``observations``. Three experiments have been performed. In  the first experiment we allow   all $\alpha$ to depend on the coordinate on the boundary; in the second experiment it is  $\alpha_0$ only that depends on coordinate, and in the third experiment all  $\alpha$ are independent. Assimilations was  performed minimizing the 4D-VAR cost function  $\costfun_4$ \rf{costfn-4} with assimilation window  $T=50$ days and the results are shown in \rfg{unif}. The assimilation window is delimited by the vertical dashed line in the figure.

\begin{figure}
  \begin{center}
  \begin{minipage}[r]{1\textwidth} 
  \centerline{\includegraphics[angle=0,width=0.49\textwidth]{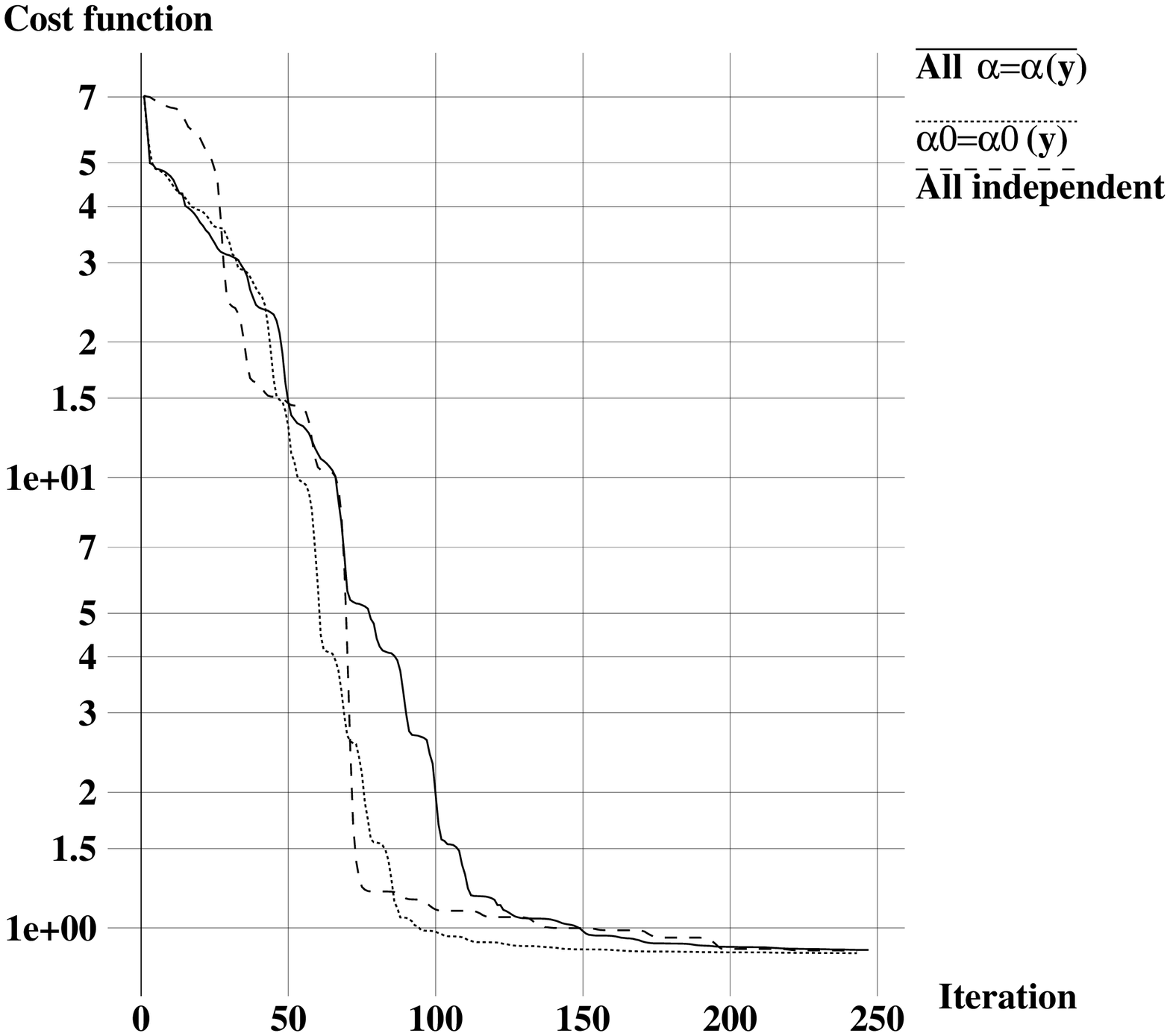}
              \includegraphics[angle=0,width=0.49\textwidth]{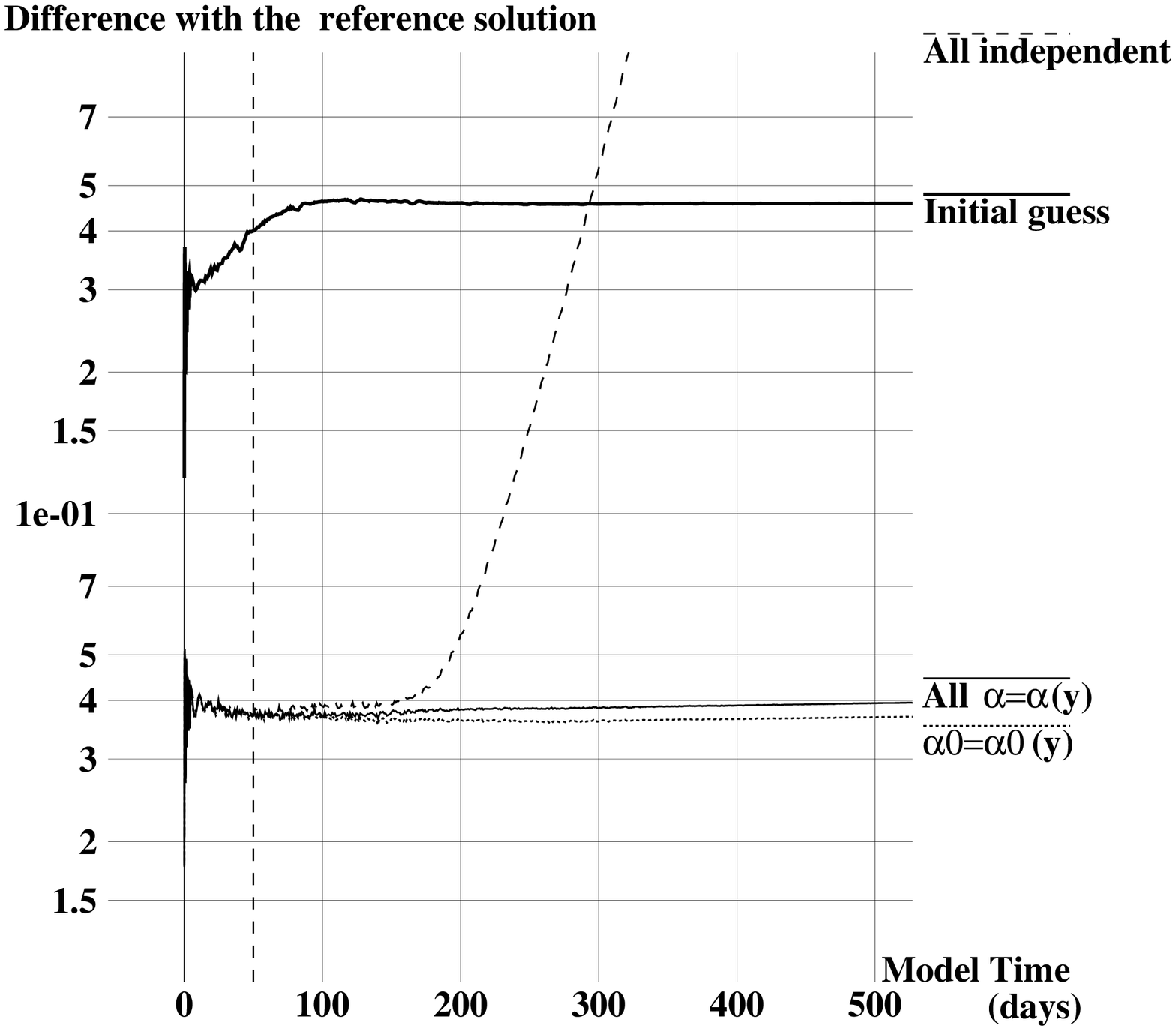}}
  \caption{Convergence of the cost function in the minimization procedure (left) and evolution  of the norm of difference with the reference solution (right)  in the experiments with all coefficients $\alpha$ particular for each point (solid line), with just $\alpha_0$ depending on the coordinate (small dashes) and all uniform $\alpha$ (long dashes).   }
  \end{minipage} 
  \end{center} 
\refstepcounter{fig}
\label{unif}
\end{figure}

As one can see in \rfg{unif}, 150 iterations are sufficient for the minimization process to converge in general. In all subsequent iterations the value of the cost function is modified just a little and can be considered as already converged to the same value in all three experiments. We can notice only a slightly different convergence rate. When the coefficients $\alpha$ are particular for each  grid point, the convergence is a little slower. The procedure takes about 150 iterations to achieve the value obtained after 90 iteration when the control coefficients are uniform or partly uniform (just $\alpha_0$ depends on the coordinate). The final value of the cost function is 80 times lower than the starting value, that means the solution is much improved in the assimilation process.

However, the particularity of the data assimilation applied to control the discretization of operators consists in the fact that obtained numerical approximation of derivatives and interpolations can be instable. In the assimilation window this numerical instability  may be dumped by the minimization, but it may bring the solution to infinity beyond the  window.

One can observe this phenomenon in the \rfg{unif} on the right where the evolution of the norm of the difference $\xi(t)$ \rf{xi} is plotted during 500 days (the assimilation window, as it has been already mentioned, is 50 days long). The solution obtained by the model with uniform coefficients $\alpha$ is indeed instable. In the assimilation window the line that corresponds to this solution is indistinguishable from two other lines, but, after the assimilation's end, the difference grows up and leave the picture's frame. 

Hence, as it has been shown above, using all uniform coefficients is not the best choice. Mass and momentum fluxes can not be balanced on the boundary and the assimilation procedure have to look for the optimum in the domain where the scheme is instable. Consequently, at least $\alpha_0$ should be allowed to be variable and specific in each point near the boundary in order to allow the integral flux across the boundary to be balanced.  This requirement substantially increases the dimension of the control space. If all control coefficients do not depend on the coordinate,  the number of coefficients for the linear shallow water model may be as few as 48. Requiring the dependence of $\alpha_0$ on the position of the boundary point, we increase the dimension of the control space up to $32+16\tm N$, where $N$ is the number of grid nodes on the boundary part where the control is performed. In the experiment with the coarse resolution ($h=133$ km, $N=30$), for example, the dimension increases from 48 to 512. This may be really penalizing when we try to avoid the development of the adjoint model and  calculate the gradient of the cost function by a finite difference method. However, even in this case the number of coefficients to control is much less than the dimension of the model's state (900 points for each of 3 variables). Moreover, if we refine the grid,  the number of coefficients will always be less than number of model's variables because the number of $\alpha$ is  proportional to the length of the boundary, while the dimension of the model state is proportional to the area of the domain. 

So far, the difference between experiments with all variable coefficients and with variable $\alpha_0$ only is low, we shall use this configuration in all experiments below. That means, all  the approximations near boundary will be calculated by the formula
\beq
(D_x u)_{m-1/2,j-1/2}=\fr{1}{h}\biggl((\alpha_{0,m}^{D_xu})_j+\sum\limits_{k=1}^{K} \alpha^{D_xu}_{k,m} u_{k-1/2,j-1/2}\biggr) \mbox{ for } m=1,2,\ldots, M \label{bndschj}
\eeq

The set of optimal coefficients $\alpha$ obtained in this assimilation is the following. All coefficients $\alpha$ other than $\alpha_0$ have moved very little from their initial guesses (less than 0.01), except $\alpha^{S_x\eta}$ on the left boundary. The approximation of the interpolation $S_x \eta$ become 
\beq(S_x \eta)_{1,j-1/2}=(\alpha_0^{S_x\eta})_j+  0.46 \eta_{1/2,j-1/2} + 0.52 \eta_{3/2,j-1/2}
\label{sxp}
\eeq
 On the other hand, coefficients $\alpha_0$ have been really used as controls in this experiment. Their deviation from the initial guess (zero) is non negligible and, indeed, they  depend  on the coordinate. Three of them, that have been obtained for the operators in $x$ ( $(\alpha_{0,1}^{D_xu})_j,\;  (\alpha_{0,1}^{S_xu})_j$ and $(\alpha_0^{S_x\eta})_j$) are shown in \rfg{unif-alp} on the left, and three others, for the operators in $y$ ($(\alpha_{0,1}^{D_yv})_i,\;  (\alpha_{0,1}^{S_yv})_i$ and $(\alpha_0^{S_y\eta})_i$) on the right. 

\begin{figure}
  \begin{center}
  \begin{minipage}[r]{1\textwidth} 
  \centerline{\includegraphics[angle=0,width=0.49\textwidth]{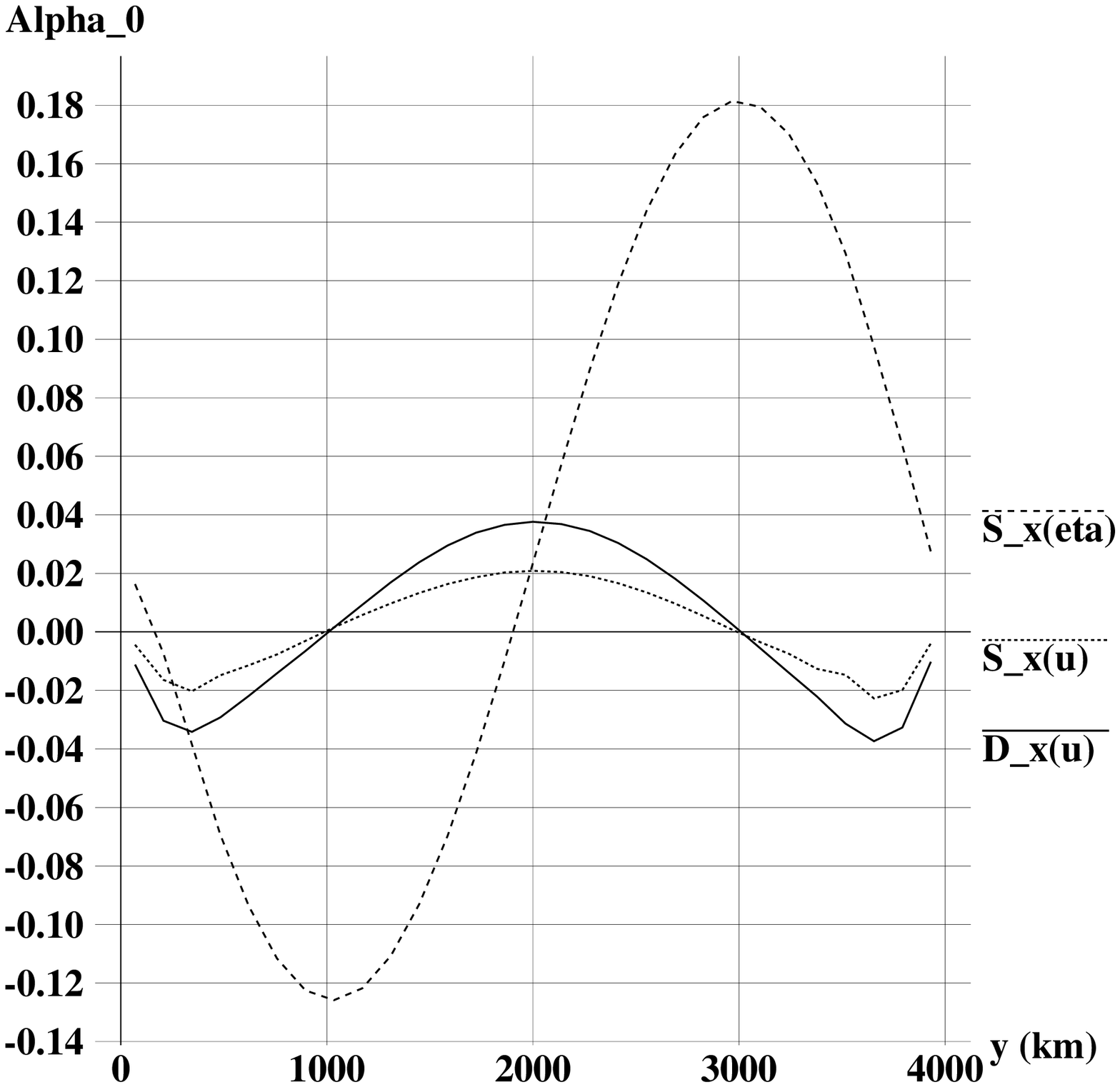}
              \includegraphics[angle=0,width=0.49\textwidth]{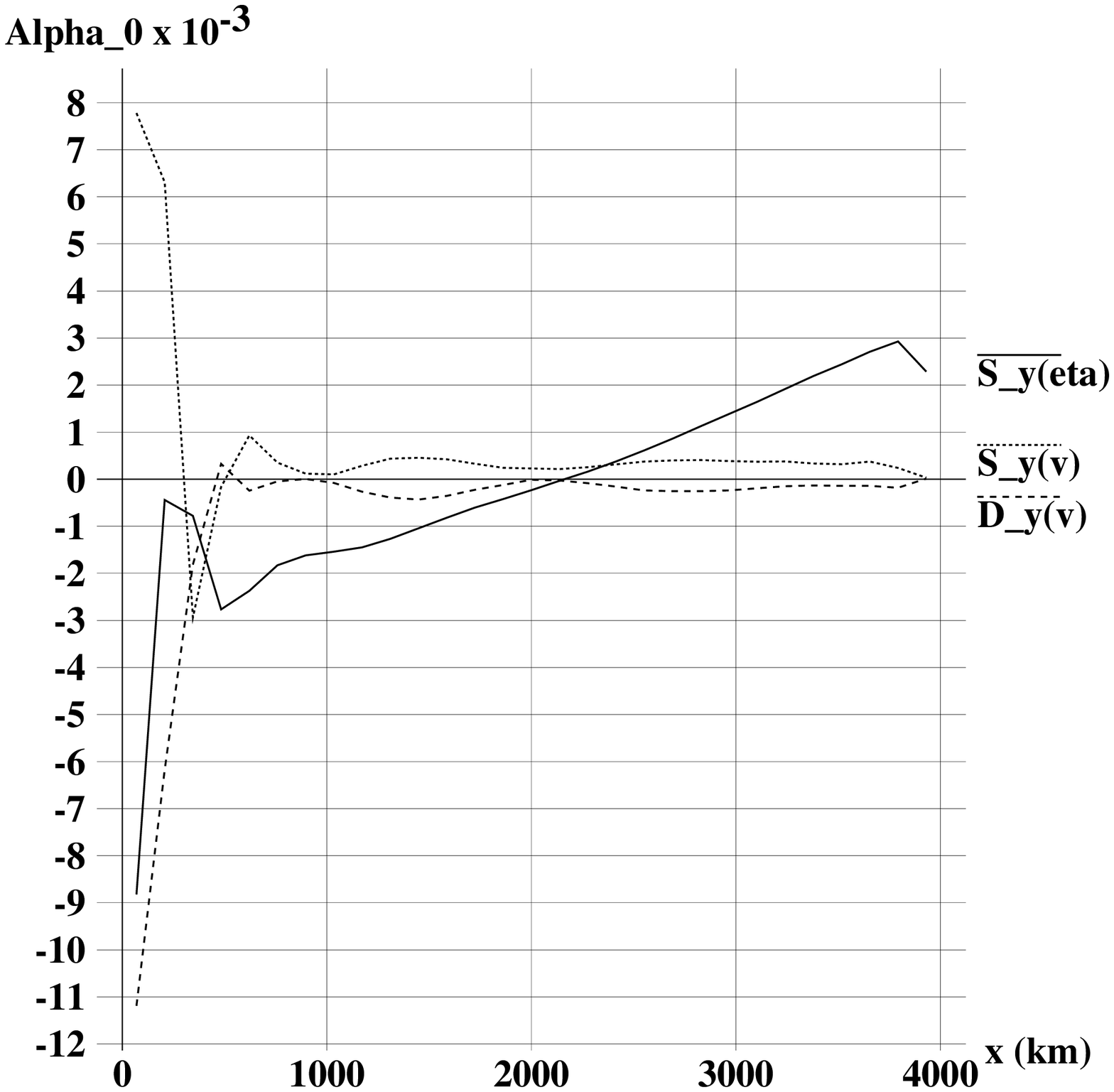}}
  \caption{Optimal coefficients $\alpha_0$ for operators in $x$ (left) and operators in $y$ (right) as  functions of $y=jh$ and $x=ih$. }
  \end{minipage} 
  \end{center} 
\refstepcounter{fig}
\label{unif-alp}
\end{figure}
On the left part of this figure we can see two important phenomena. The first one consists in the fact that the flux across the boundary is  balanced. Despite no balance requirement has been imposed in the data assimilation procedure, the integral of the  flux over the boundary is close to zero. The normal component of the velocity  on the left boundary (i.e. $u$ component) shows that the flux leaves the domain on the North and on the South of the boundary and returns back in the middle. 

The second phenomenon consists in the fact that coefficients $(\alpha_{0,1}^{D_xu})_j$ and $(\alpha_{0,1}^{S_xu})_j$ have a certain similarity.  As it has been already mentioned, they have been considered as completely independent on each other,  but the result shows the coefficients $  (\alpha_{0,1}^{S_xu})_j$  are very close to one half of  $(\alpha_{0,1}^{D_xu})_j$.

Analyzing  coefficients $\alpha_0$ for operators in  $y$ shown in \rfg{unif-alp} on the right, we see  their values are much smaller than for operators in $x$. It is reasonable because there is no boundary layers near the North and near the South boundaries. These coefficients have maxima only in the  North-Western and the South-Western corners of the domain  because  of influence of the western boundary layer. Out of this layer, only coefficients $\alpha_0^{S_y\eta}$ (solid line) show significant difference from zero. The operator 
$S_y\eta$ is applied to the result of the interpolation $S_x u$.   That means the interpolation of the velocity $u$ to points $v$ is performed by formula
$$
u_{i+1/2,1}=(\alpha_0^{S_y\eta})_i+\fr{u_{i+1,3/2}+u_{i,3/2}+u_{i,1/2}+u_{i+1,1/2}}{4}
$$
which differs from the standard second order interpolation only by adding $(\alpha_0^{S_y\eta})_i$ that represents a flux to the West in the western half of the domain and the flux to the East in the eastern half as it is  shown in \rfg{unif-alp} on the right.

The next question we address in this section concerns  the length of the assimilation window and the form of the cost function. A priori, we know the procedure is computationally less efficient when $T$ is long. Computing time per iteration increases with $T$ because each iteration is composed by the integration of the direct model from 0 to  $T$ and corresponding backward  integration of the adjoint model. Consequently, it would be more computationally efficient to choose a small $T$. However, it is the precision of the assimilation that we want to improve and the precision must be used as criterium in the choice of the assimilation window.

The choice of the cost function among $\costfun_4,\; \costfun_{fp}$ and $\costfun_T$ is not obvious also. As it has been already noted, the cost function $\costfun_4$  gives the same importance to all part of the assimilation window, while $\costfun_T$ and $ \costfun_{fp}$ consider the end of the interval to be more important than the beginning. 

In order to see the influence of the form of the cost function and the assimilation window, we perform the  set of experiments with different cost functions and different assimilation windows. In all experiments we stop the minimization process at different stages of convergence: in the very beginning, just after 10 iterations; in the middle after 50 iterations and after 200 iteration allowing almost complete convergence. So far, final values of different  cost functions are difficult to compare  among them, we shall compare final values of $\xi$ \rf{xi} that is independent on the cost function in use. Moreover, in each experiment we shall compare not only $\xi(T)$ calculated at the end of the assimilation window, but also $\xi(1000 \mbox{ days})$ in order to distinguish those assimilations that result in an instable scheme.   
 
\tabcap{\textwidth}
\begin{table}
\caption{ Values of $\xi$ obtained by assimilation with 4 different assimilation windows and 3 different cost functions.}
\begin{center}
\begin{small}
\begin{tabular}{|c|c|ccc|ccc|}
\hline
T      &Number of & \multicolumn{3}{|c|}{$\xi(T)$}&\multicolumn{3}{|c|}{$\xi(1000 $ days)}\\
(days) &Iterations&using $\costfun_{fp}$&using $\costfun_4$&using $\costfun_T$&using $\costfun_{fp}$&using $\costfun_4$&using $\costfun_T$\\
\hline
0.2& 10   &0.042 &0.042 &0.042    &$\infty$ &$\infty$ &$\infty$  \\
0.2& 50   &0.041 &0.041 &0.041    &$\infty$ &$\infty$ &$\infty$  \\
0.2& 200  &0.040 &0.040 &0.040    &$\infty$ &$\infty$ & $\infty$ \\
\hline
2& 10     & 0.12 &0.041  &0.09     &1.79  &0.74 &1.88  \\
2& 50     &0.031 &0.032 &0.032     &0.09  &0.05&0.04 \\
2& 200    &0.029 &0.030 &0.030     &1.22&$\infty$ & $\infty$ \\
\hline
20& 10    & 0.27 &0.26  &0.27	  &0.56  &0.54  &0.64  \\
20& 50    & 0.11 &0.064 &0.08	  &0.24  &0.17  &0.14  \\
20& 200   & 0.04 &0.037 &0.04	  &0.050  &0.040 & 0.036 \\
\hline
200& 10   &0.25 &0.25 &0.25	  &0.25  &0.25 &0.25  \\
200& 50   &0.21 &0.21 &0.21	  &0.21  &0.20 & 0.19 \\
200& 200  &0.05 &0.04 &0.03	  &0.049  &0.039 & 0.035 \\
\hline
\end{tabular}
\label{tab1}
\end{small}
\end{center}
\end{table}

Analyzing the Table \ref{tab1}, we can note several interesting particularities. First of all, when assimilation window is short, the convergence is very rapid. In fact, 10 iterations are  sufficient for the minimization to converge when $T=0.2$ days (just two time steps of the model). Taking into account that iterations with a so short $T$ are extremely cheap in computer time, it seems to be advantageous to perform assimilations with short windows. However, these assimilations result in an unstable scheme. Two time steps is too short interval to develop the instability.  The cost function  contain no information about stability of the scheme, consequently, the minimization chooses a set of coefficients $\alpha$ with no requirements about stability of the obtained scheme. 

If we consider assimilation window $T=2$ days, we see another interesting particularity. Ten iterations are no longer sufficient for the minimization to converge, but after 50 iterations the minimization has almost reached the minimum. Next 150 iterations allow just to reduce the cost function by $0.002$, i.e. less than 10\%. 
But these 150 supplementary iterations transform a stable  scheme  to an instable one.  After 50 iterations the scheme was perfectly stable allowing the model to be integrated  as long as necessary. At the end of 1000 days model's run we get rather low difference with the reference solution of the model on a fine grid. But continuing minimization we win a little in the cost function loosing the stability. Consequently, we should assume the optimal discretization of operators near the boundary  is situated not far from the stability limit. Performing more iterations than some maximal allowed quantity, we can cross the stability limit and obtain an instable scheme. This maximal  number of iterations depends on the assimilation window. The scheme becomes instable just after  several iterations when $T=0.2$ days because there is very few information in the cost function about stability. When $T=2$ days, the cost function knows more about stability (the solution must not diverge during at least $2$ days), and the minimization supports several dozens of iterations before become instable. The stability is not guaranteed even with $T=20$ days. This experience is not shown in the Table \ref{tab1}, but after 1000 iterations we obtain an instable scheme also. 

The conclusion, hence, is simple: short $T$ ensures a rapid convergence, but long $T$ is necessary to ensure the stability. Reasonable strategy of data assimilation in this circumstances may consist in using variable assimilation window. Performing few iterations  with short $T$ we obtain the set of coefficients $\alpha$ that ensures low cost function value. This set is corrected by assimilation with longer $T$ in order to ensure the stability of the scheme. Of course, it is important to restrict the number of iterations with short windows. Otherwise, assimilation with long windows will not be able to correct instabilities of the scheme. 

Looking at the results obtained using different cost functions  in the assimilation procedure, we see just a little difference in  final values of  $\xi(1000$ days). However, the final  $\xi$  obtained with $\costfun_T$ is a little lower than with any other cost function. Consequently, the hypothesis that the end of the assimilation window should be more heavily weighted is reasonable.  Excessive weighting of the end of the window by $\costfun_{fp}$, however, provides worse results. When we use just the final point in the cost function, the value of $\xi(1000$ days) is bigger than in all other experiments. 

Taking into account the results obtained in  this section, we shall use the cost function $\costfun_T$ and perform the  data assimilation with variable assimilation window $T$ in all experiments below. A comparison of the convergence of the cost function in experiments with fixed assimilation windows of different length and with variable windows is shown in \rfg{seq}. 
So far, the value of the  cost functions depends explicitly on $T$, we  normalize it and plot the value $\fr{2\costfun_T}{T^2}$ which depends on $T$ only implicitly.  

In this figure one can see the same phenomenon as in the Table \ref{tab1}. When the assimilation window is short, the convergence is  rapid. Just several iterations are sufficient for minimization to converge when $T=0.2$ days or $T=2$ days. Unfortunately, as we have seen in the Table \ref{tab1}, obtained numerical schemes are instable. When the assimilation window is long, more iterations are necessary to converge (about 100 when $T=20$ days and about 200 with $T=200$ days). Moreover, each of these iterations is more expensive in computer time. 

One can see in \rfg{seq} also, the convergence of   $\fr{2\costfun_T}{T^2}$ with the variable window is rapid, the final value of the cost function is the same and obtained set of coefficients $\alpha$ provide a stable scheme. In the experiment with variable window we have performed 4 iterations with $T=0.2$ days, 8 iterations with $T=2$ days and 20 iterations with each of $T=20$ and 200 days. In fact the computing time  in the experiment with variable window is equivalent to 23 iterations with  $T=200$ days, that means we obtain the same result almost 10 times more rapidly in terms of CPU time.

\begin{figure}
  \begin{center}
  \begin{minipage}[r]{1\textwidth} 
  \centerline{\includegraphics[angle=0,width=0.79\textwidth]{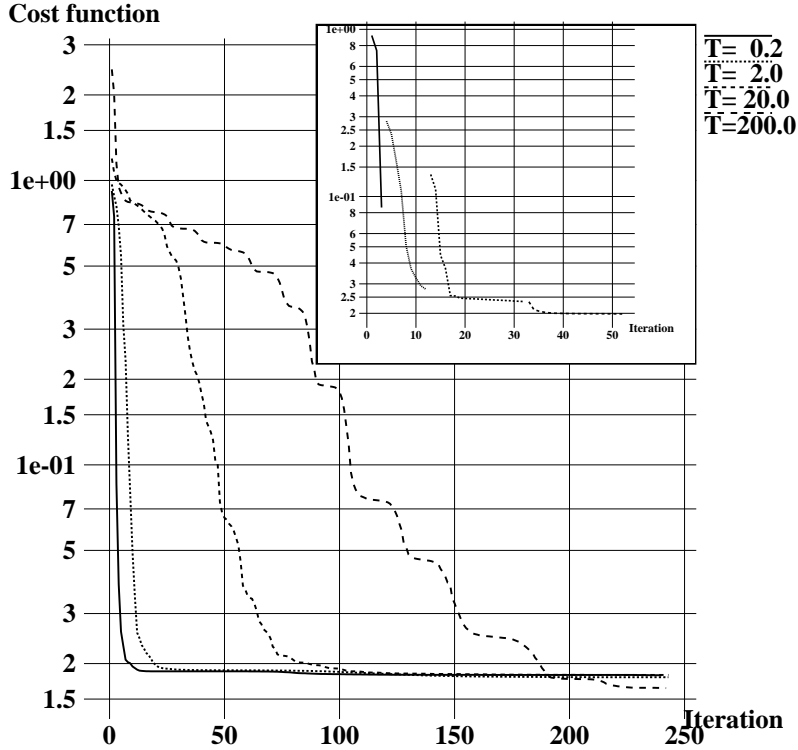}}
  \caption{ Convergence of the normalized cost function in the assimilation procedure with different assimilation windows and with sequentially increasing $T$. }
  \end{minipage} 
  \end{center} 
\refstepcounter{fig}
\label{seq}
\end{figure}

\subsection{Assimilation window and the number of control parameters. }

As it has been already noted, the difference with controlling boundary conditions consists in the liberty to increase the number of controlled parameters $\alpha$ increasing either $K$ or $M$ (or both) in \rf{bndschj}. In this section we perform  a set of experiments with different $K$ or $M$  looking for  the solution of the model on the coarse grid that is close to the reference model's solution on a fine grid. To measure the distance between two model's solutions we use the norm of the difference $\xi$ \rf{xi} between them taken at 1000th  day of the model time. The time  1000 days is chosen as sufficiently long time at which all the intermittent processes are over and the model has reached its equilibrium.  (the model is linear and dissipative, so the equilibrium exists and represents the global attractor of dimension 0). From the Table \ref{tab1} we see that  with $K=2$ and $M=1$  we can obtain $\xi(1000$ days)=0.035 while the value of $\xi(1000$ days) calculated for the original coarse resolution model with classical discretizations of all operators is equal to 0.46. 

That means, performing data assimilation,  we have obtained the solution that is more than 10 times closer to the reference one than the initial guess. In fact, optimal discretization of operators near boundary has allowed us to obtain the solution  on the coarse grid (grid step $h=133$ km) that is similar to the solution on the medium grid ($h=44$ km). Indeed, the difference $\xi$ between equilibria on the medium and on the fine grids  is equal to 0.037 that is almost the same as for the coarse grid's solution with optimal $\alpha$.   The profile of the velocity $v(x,y)$ with $x=1000$ km of  the solution  on the medium grid is presented in \rfg{profil} together with coarse grid's and fine grid's solutions. We see that there is no spurious numerical oscillations in the solution on the medium grid. Hence, the data assimilation helped us to remove these oscillations on the coarse grid also. 

However, increasing  $K$ and $M$ does not help to diminish the difference with the reference model. Several experiment have been performed with $M=2$ and $M=3$ and with $K$ varying from 2 to 7, but  no significant improvement has achieved. The lowest  value of $\xi$ obtained in these experiments is equal to 0.012. This value  is slightly lower than 0.035 obtained with  $K=2$ and $M=1$, but these values  are  of the same order. 

That means controlling the discretization of operators near boundary we can avoid spurious numerical oscillation due to under-resolved boundary layer and  obtain the solution similar to the solution  on the 3 times finer grid. But we can not go further and to get a solution  closer to the reference one.  This phenomenon can be explained by the analysis of the error  field, i.e. the differences $(u_{i,j}- u^{obs}_{i,j}),\; (v_{i,j}- v^{obs}_{i,j}) $ and $(\eta_{i,j}- \eta^{obs}_{i,j})$ that are used to calculate $\xi$. The error in the initial guess is concentrated in the boundary layer while the model with optimal discretization near the boundary provides a solution with more uniform error's distribution in the whole domain. Thus, the error in the  sea surface elevation in the initial guess  reaches 30 meters in the boundary layer and 2-3 meters in the eastern part of the square.  But after data assimilation we get an error of  1 meter near the western boundary and 0.3 meter in the eastern part. More uniform distribution of the difference with the reference solution indicates that the error produced by numerical  scheme in the interior of the domain become relatively more important. And so far, we do not control the interior scheme we can not reduce this error. No matter how many $\alpha$ we use to control the approximation near the  boundary, they can not improve the approximation in regions far from the boundary. 

In order to verify this hypothesis, we perform a similar set of experiments with the fourth order approximation's schemes in all internal points of the domain. As well as before, all operators have been approximated by \rf{intrnlsch}, but with coefficients  $ a^D_k=\fr{1}{24h}(1,-27,27,-1)$  for $ k=-2,-1,0,1$ for derivatives and  $ a^S_k=\fr{1}{8}(-1,9,9,-1)$  for interpolations. The same initial guess  $\alpha$ as in experiments with the  second order approximation near the boundary has been used.

Data assimilation in all these experiments has been performed with variable windows from $T=2$ to $T=200$ days.  As we have seen before, obtained numerical scheme may be instable and generate a rapidly divergent  solution after the end of assimilation. For this reason, together with the cost function over the assimilation window $T$ we look at the value of the norm of the difference between the solution on the current iteration and the reference solution $\xi$ \rf{xi} taken at the end of the test interval, i.e. on the  1000th day (i.e. 5 times the longest assimilation window).  The minimization process  is interrupted  when the scheme begins to loose its stability. That means the value of $\xi$ taken at the end of the test interval $T=1000$ days  starts to grow despite decreasing of $\xi$ at the end of the assimilation window.  
The lowest value $\xi(1000$ days) obtained in the assimilation  is shown in the Table \ref{tab2}. 

Considering the initial guess, we can see, that  the value of $\xi$ at the end of test interval is bigger for the model with the fourth order scheme than with the second order one. The reason of this is clear: high order scheme work worse when principal physical scales are not resolved explicitly. Indeed,  the coarse grid of the model does not resolve the Munk boundary layer \rf{Munk}. The grid step $h$ is approximately twice  the Munk parameter $2\delta=2(\mu/\beta)^{1/3}$. But it is the ratio $\biggl(\fr{h}{2\delta}\biggr)^{n}$ where $n$ is the order of approximation  that determines the approximation error \rf{taylor}. Higher is the order of approximation, bigger is this ratio and bigger is the error in the approximation of the boundary layer. 

Another interesting phenomenon that can be seen in the Table \ref{tab2} consists in relative independence of the final value of $\xi$ on the number of control coefficients $K$. No matter how many terms participate in the linear combination that approximates the derivative or interpolates the function, the minimization procedure converges always to the same value. Obviously, this fact  is in agreement with the feature discussed above. Using more terms  in linear combination allows us to increase the approximation's order. One can construct a second order scheme with two terms only, but with 4 terms any approximation up to fourth order is possible. However, as we have seen, increasing the order of approximation does not help to improve the model on this coarse grid. Moreover, the interpolation obtained by data assimilation \rf{sxp}
 approximates the function with zero order $O(1)$. Consequently, it is useless to increase the number of terms $K$ in the linear combination because the  data assimilation does not look for increasing the order. 

On the other hand, increasing $M$ with the fourth order scheme really improves the result. Controlling numerical schemes in more than one point near the boundary allows us to obtain the solution which is very close to the reference one, obtained on the nine times finer grid. In fact, controlling $\alpha$ in  one additional point divides the final value of $\xi$ by 10. When $M$ is equal to 4, $\xi(1000$ days) is as small as $4\tm 10^{-4}$, that means the solution with optimal $\alpha$ is almost indistinguishable from the reference solution.

\tabcap{\textwidth}
\begin{table}
\caption{ Values of $\xi(1000$ days) obtained by assimilation with  different number of control coefficients by second and  fourth order schemes.}
\begin{center}
\begin{small}
\begin{tabular}{|c|cl|}
\hline
$K$ and $M$ & Second order &Fourth order\\
\hline
Initial Guess& 0.46& 1.3\\
Medium grid& 0.039& 0.028 \\
\hline
$K=2,\;M=1$  & 0.035& 0.15\\
$K=3,\;M=1$  & 0.035& 0.15\\
$K=5,\;M=1$  & 0.035& 0.15\\
\hline
$K=5,\;M=2$  & 0.018& 0.010\\
$K=7,\;M=3$  & 0.013& 0.001\\
$K=7,\;M=4$  & 0.012& 0.0004\\
\hline
\end{tabular}
\end{small}
\end{center}
\label{tab2}
\end{table}

Let us look at  the set of control coefficients $\alpha$ obtained in the last experiment: $K=7,\;M=4$  with the fourth order approximation in the interior part of the domain. It is this experiment that provides the closest trajectory to the reference one.  This set of control parameters contains more elements that the set described in the previous section. We control discretization of derivatives and interpolation operators in four points near boundary ($m=1,2,3,4)$ and each derivative and interpolation is approximated by linear combination of the function's values in seven adjacent points ($k=1,\ldots ,7)$. 

However, we get a very similar discretizations in the closest to the boundary points ($m=1$). All coefficients $\alpha$ other than $\alpha_0$ have also moved very little from their initial guess (less than 0.01), except $\alpha^{S_x\eta}$ on the left boundary. The approximation of the interpolation $S_x \eta$ become 
$$
(S_x \eta)_{1}=(\alpha_0^{S_x\eta})+  0.46 \eta_{1/2} + 0.50 \eta_{3/2}- 0.02 \eta_{5/2}- 0.006 \eta_{7/2}- 0.006 \eta_{9/2}- 0.001 \eta_{11/2}
$$
Comparing this expression with \rf{sxp}, we can see that the difference consists just in replacement of the term $0.52 \eta_{3/2} $ by the linear combination of $\eta$ in next nodes. 

Coefficients $\alpha_0$ have also  been really used as controls in this experiment and they have almost the same values as in the experiment with $K=2,\;M=1$.  Their dependence on coordinate is very similar to the dependence shown in  \rfg{unif-alp}.  

The coefficients $\alpha_{k,2}$ that are used in approximations in nodes next to the first has been modified  more significantly than coefficients $\alpha_{k,1}$ in nodes adjacent to the boundary.  The deviation from the initial guess exceeds 0.01 for all operators in the Munk Layer and for $D_yv,\; S_yv,\; S_y\eta$ near the North and the South boundaries. Coefficients $\alpha_{0,2}$ and $\alpha_{0,3}$ $\alpha_{0,4}$  in the Munk layer are shown in \rfg{unif-alp234}.

\begin{figure}
  \begin{center}
  \begin{minipage}[r]{1\textwidth} 
  \centerline{\includegraphics[angle=0,width=0.49\textwidth]{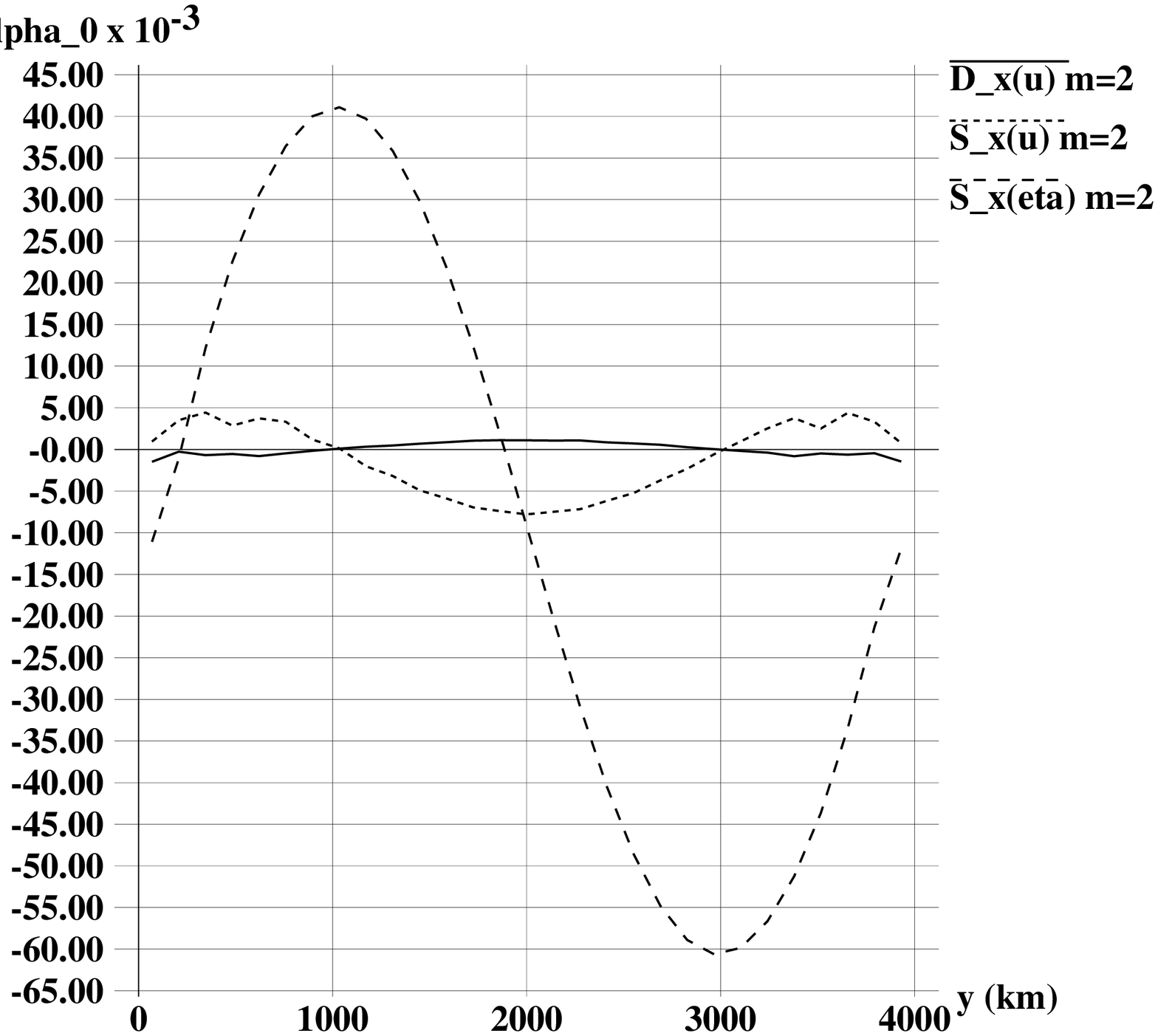}
              \includegraphics[angle=0,width=0.49\textwidth]{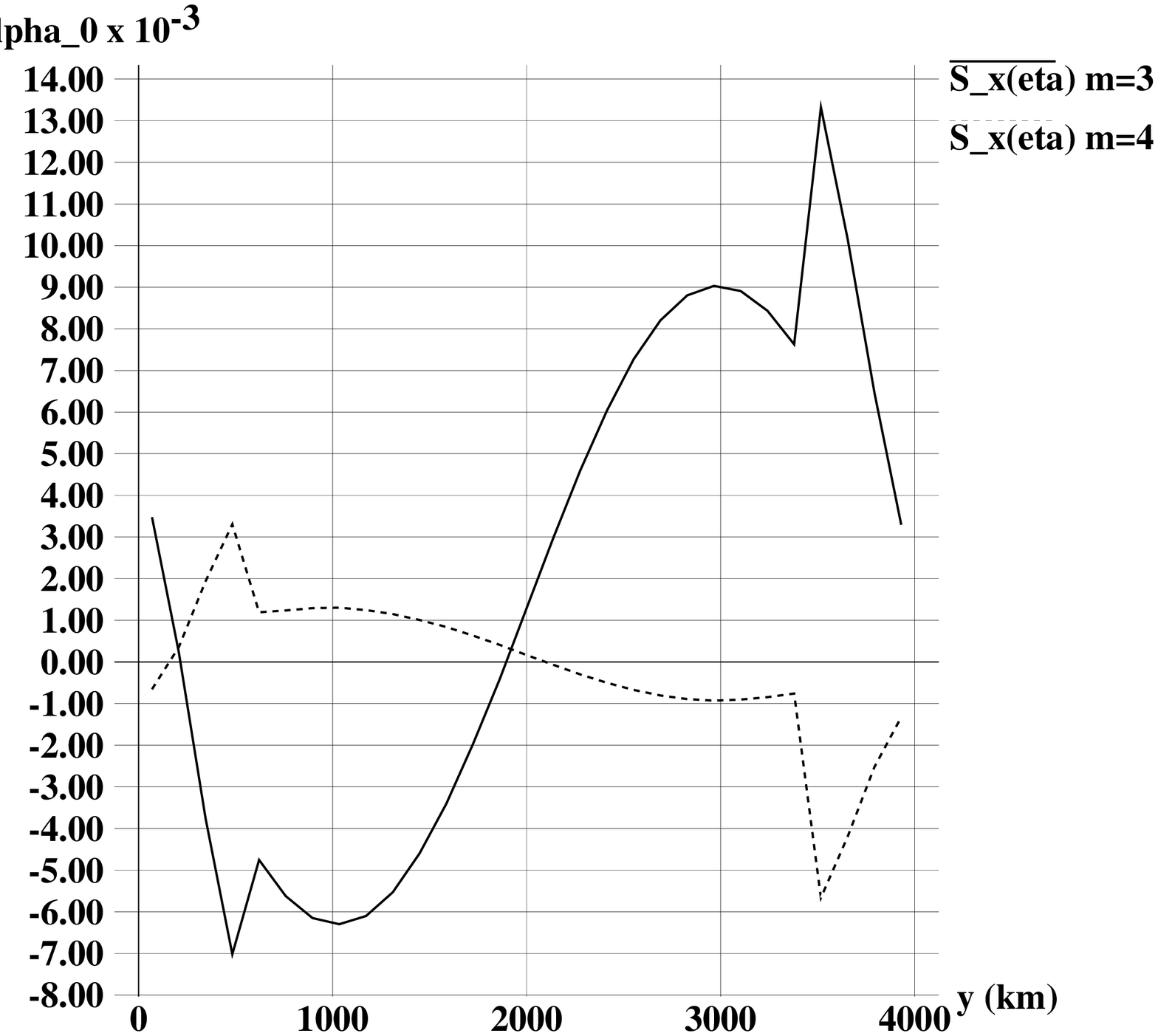}}
  \caption{Optimal coefficients $\alpha_0$ for $m=2$ (left) and $m=3$ and 4 (right) as  functions of $y=jh$. }
  \end{minipage} 
  \end{center} 
\refstepcounter{fig}
\label{unif-alp234}
\end{figure}

Comparing this figure with the \rfg{unif-alp} on the left,  we can see that all $\alpha_{0,1}^{S_xu},\; \alpha_{0,1}^{D_xu},\; \alpha_{0,2}^{S_xu}$ and $\alpha_{0,2}^{D_xu}$ are proportional to each other as well as $\alpha_{0,1}^{S_x\eta}$ and $\alpha_{0,2}^{S_x\eta}$. The data used to plot these figures allows us to show  that  $\alpha_{0,2}^{S_xu}=-\fr{\alpha_{0,1}^{S_xu}}{2.7}$ and   $\alpha_{0,2}^{D_xu}= \fr{\alpha_{0,1}^{D_xu}}{34}$. 

Concerning operator $S_x\eta$, we see that coefficients  $\alpha_{0,1}^{S_x\eta}$  and $\alpha_{0,2}^{S_x\eta}$ may reach values 0.18  and 0.04 respectively. Taking into account the fact that these are dimensional variables, that means additional elevation up to 18 and 4 centimeters is added to the interpolation of the sea surface hight near boundary.

\subsection{Using restricted observational data.} 

In the previous section we have supposed that observational data are available for all variables. However,  it is the altimetry that is the most probable candidate to be an observable variable in real data assimilations. In this section we suppose to have the sea surface elevation $\eta$  as the only observable data. As well as before, these data are produced by a high resolution model on the 9 times finer grid, but we put $w_u=w_v=0$ in \rf{sp} in order to neglect all the information about $u$ and $v$ in the cost function and in its gradient. 

However, assimilating the sea surface elevation only, we hope to be capable to identify numerical schemes for all variables. Consequently, comparing the quality of assimilation we keep former $w_u$ and $w_v$ in \rf{xi}. Thus, the norm of the difference $\xi$ between the solution of the reference model and the solution with optimal numerical  scheme on the boundary takes into account the information about all variables. The difference $\xi$ is, consequently, directly comparable with corresponding $\xi$ obtained  in previous experiments. 

Several experiments with different $K$ and $M$ have been performed in the same way as in the previous section. We are also using $\xi(1000$ days)  as the final measure of data assimilation's quality because this value, being taken well beyond the assimilation window, ensures the stability of the scheme.  Results of the assimilation of altimetry data only are shown in the Table \ref{tab3} together with several results of assimilation of all variables already shown in the  Table \ref{tab2}. 

\tabcap{\textwidth}
\begin{table}
\caption{ Values of $\xi(1000$ days) obtained by assimilation of all 3 variables and of the altimetry data only. }
\begin{center}
\begin{small}
\begin{tabular}{|c|clc|clc|}
\hline
&\multicolumn{3}{|c|}{All variables}&\multicolumn{3}{|c|}{Only $\eta$}\\
$K$ and $M$ & 2nd order &4h order&Iterations& 2nd order &4th order&Iterations \\
\hline
Initial Guess& 0.46& 1.3 & & 0.46 &1.3& \\
\hline
$K=2,\;M=1$  & 0.035& 0.15  & 200 &0.042 & 0.18 &1 200\\
$K=5,\;M=2$  & 0.016& 0.010 & 1 000&0.016 & 0.015&5 000\\
$K=8,\;M=4$  & 0.012& 0.0004& 3 500&0.016 & 0.0006&15 000 \\
\hline
\end{tabular}
\end{small}
\end{center}
\label{tab3}
\end{table}
Analyzing the Table \ref{tab2}, we have concentrated our attention on the final results of the assimilation and namely on the value of the difference with the reference solution. Now, when we restrict the number of observable variables, we should also pay attention to the cost of assimilation. As it has been already mentioned, results in the Table \ref{tab2} have been obtained by data assimilation with variable window $T$. To create the Table \ref{tab3}, we have re-obtained the same results assimilating data with a fixed window $T=200$ days. The minimization procedure in each experiment  has been stopped after stabilization of  the cost function.  The number of iterations in this case is proportional to the computer time spent to minimize the cost function. This value is, consequently, comparable in different assimilation experiments. 

Comparing the number of iterations shown in the Table \ref{tab3}, we see a valuable growth of  computing time necessary to converge the minimization process. The number of iterations increases rapidly with increasing of the number of control parameters. Bigger $K$ and $M$ ensures lower difference with the reference solution, but the assimilation becomes more expensive. In fact, the convergence rate (i.e. average decrease of the cost function per iteration) is independent on the number of control parameters, but more iterations are necessary to reach lower final value of the cost function. Thus, intermediate values of $\xi(1000$ days) obtained after  200 and 1000 iterations in the experiment with $K=8,\;M=4$ are equal to 0.15 and 0.01 respectively, that corresponds exactly to the final values of $\xi$ in the experiments with $K=2,\;M=1$ and $K=5,\;M=2$.  

 Restricted number of observable variables leads also to slower convergence. The number of iterations  with only one observable variable $\eta$ is approximately 5-8 times higher than in experiments assimilating data of all three variables.  This is not surprising:  spurious oscillations due to under resolved Munk layer are the most visible in the velocity $v$ while the sea surface height data are assimilated. The information, hence,  must be transferred to $v$.  This takes more time and  requires supplementary iterations. 

Concerning assimilation results with only one observable variable, we can see that  final values of $\xi$ are close to corresponding values obtained assimilating data collected for all variables. 
Values in the left part of the Table \ref{tab3} are a little lower, but this seems to be due to preliminary stop of the minimization procedure.  Convergence becomes so slow that matches the stop condition despite further decrease is still possible. 

The hypothesis of the preliminary stop is also confirmed by the analysis of obtained assimilation results. As well as in previous experiments, all $\alpha$ except $\alpha_0$ have moved very little from their initial positions. Coefficients $\alpha_0$ have been subjected to  more significant modifications. Their  values for operators acting in $x$ coordinate in the experiment with $K=8,\; M=4$ are shown in \rfg{unif-alp-h} for $m=1$ and 2. This figure can be compared with 
\rfg{unif-alp} and \rfg{unif-alp234}. We can see that $\alpha_{0,1}$ for all operators $D_xu,\; S_xu$ and $S_x\eta$ have converged to  very similar values as in previous  experiments.   Preliminary stop of the minimization procedure results in irregularity of $\alpha_{0,2}$ for operators $D_xu$ and $ S_xu$. Their lines have the same  appearance as corresponding lines in \rfg{unif-alp234}, but they contain  w some irregular noise which may represent the consequence of too early stop of the minimizator.

\begin{figure}
  \begin{center}
  \begin{minipage}[r]{1\textwidth} 
  \centerline{\includegraphics[angle=0,width=0.49\textwidth]{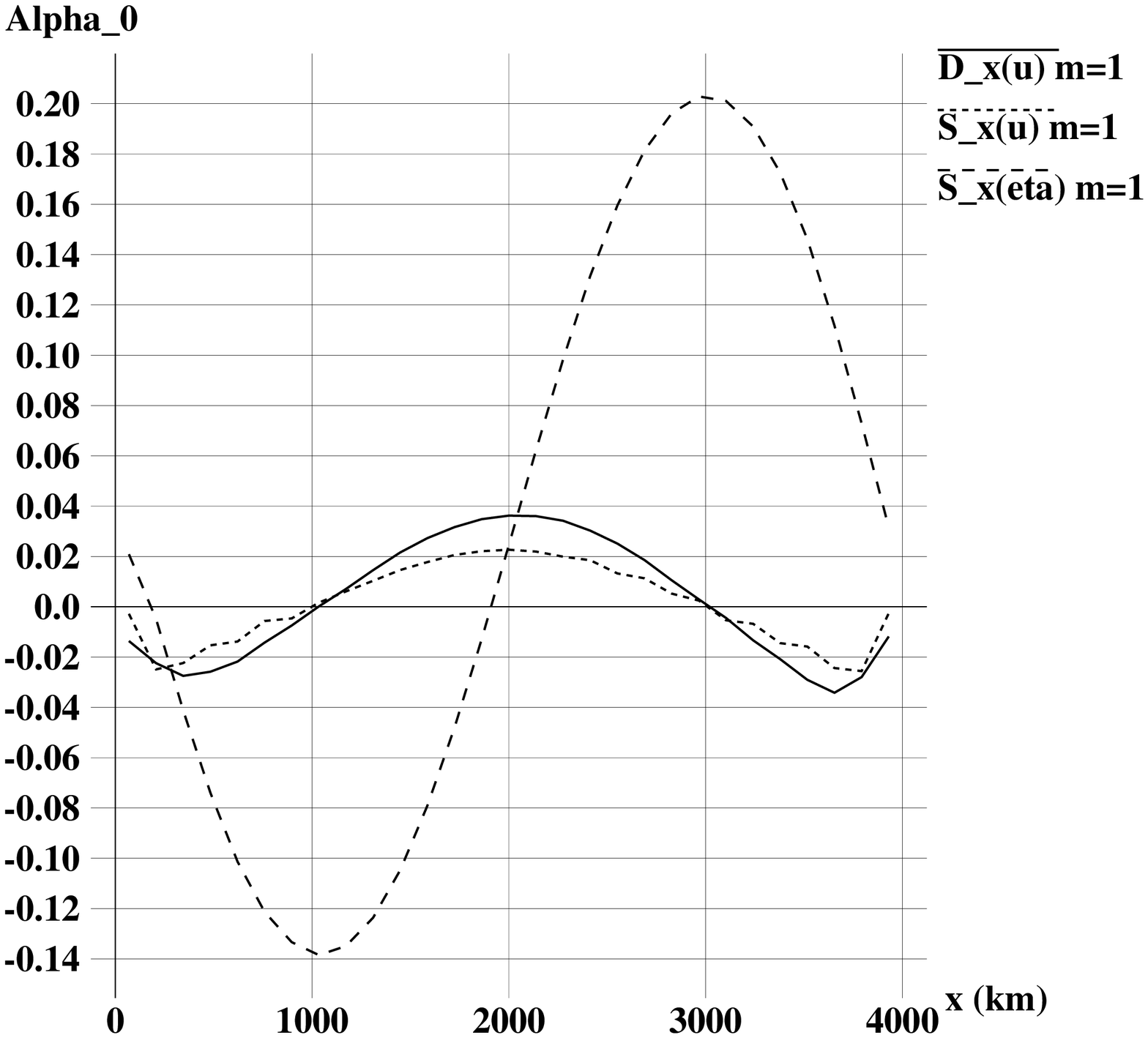}
              \includegraphics[angle=0,width=0.49\textwidth]{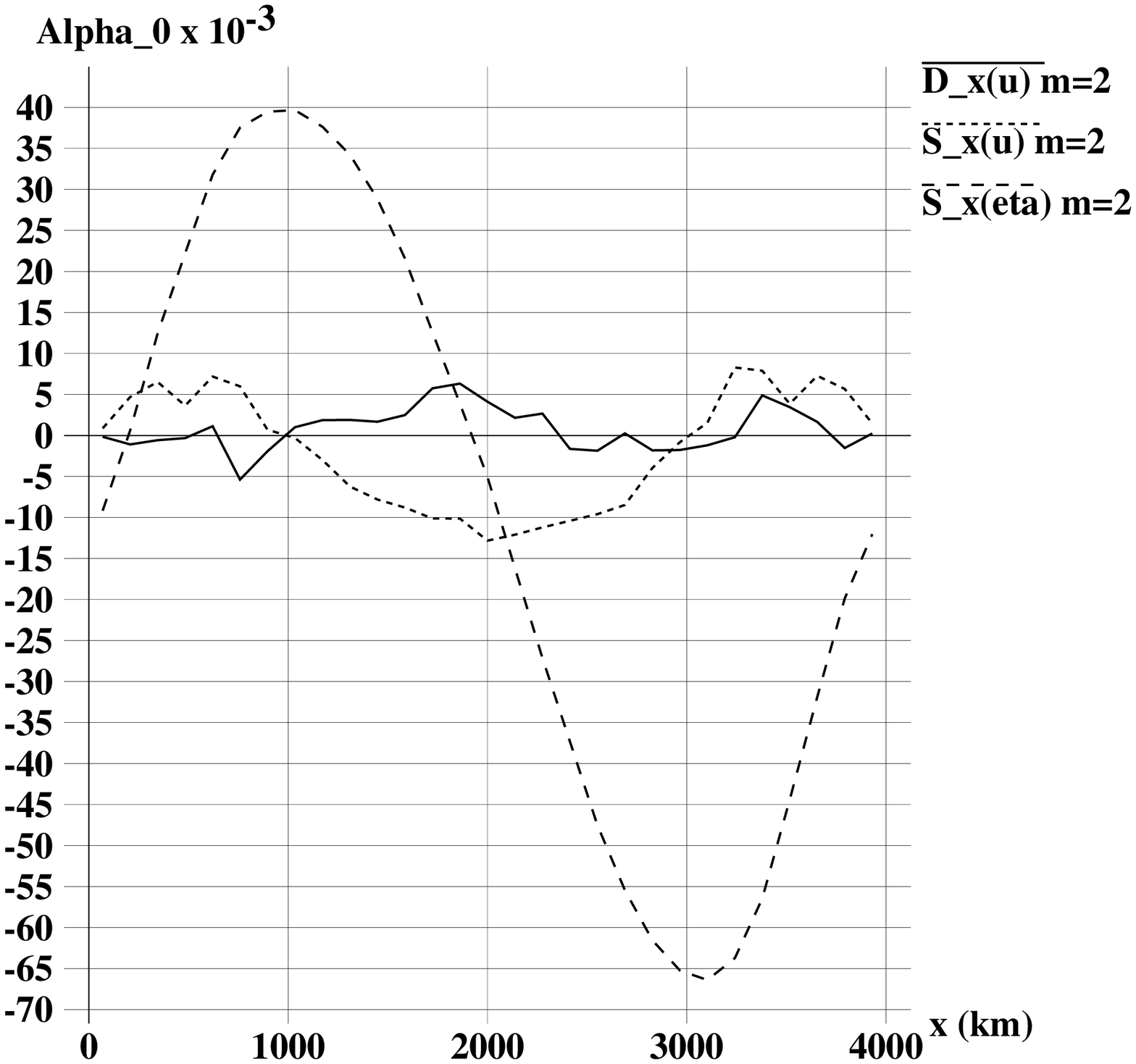}}
  \caption{Optimal coefficients $\alpha_0$ for $m=1$ (left) and $m=2$ (right) as  functions of $y=ih$ in the experiment with the only observable variable $\eta$. }
  \end{minipage} 
  \end{center} 
\refstepcounter{fig}
\label{unif-alp-h}
\end{figure}

Consequently, the lack of observations leads to a slower convergence. Assimilating the sea surface elevation only, we must take into account that the computational cost of the procedure  may be multiplied by 10.

\section{Waves}

Forced and dissipative linear shallow water  model  possesses a stationary solution that was considered in the previous section. So far, the numerical scheme under control does not depend on time, the use of a stationary solution represent an obvious advantage. Of course, it is easier to find optimal discretization when the flow is stable. 

In this section we shall look for the set of $\alpha$  that is optimal for a non-stationary solution of the shallow water model and namely inertia gravity and Rossby waves. The shallow water  model  is written without forcing nor dissipation:  

 \beqr
\der{u}{t} &=& (f_0+\beta y) S_x S_y v-  g G_x\eta 
\nonumber \\
\der{v}{t} &=&- (f_0+\beta y) S_y S_x u - g G_y \eta \label{swwv-grid} 
 \\
\der{\eta}{t} &=& - H_0(D_x u+D_y v) \nonumber
\eeqr

The system becomes hyperbolic with exact conservation of total energy and momentum. The mass is also conserved as well as in the forced-dissipative model. 
So far, there is no more harmonic dissipation in the problem,  boundary conditions have been reformulated in order to ensure the existence and uniqueness of a solution:
\beq u(0,y)=u(L,y)=v(x,0)=v(x,L)=0 \label{bcwaves}\eeq

We keep the same model's parameters as above, including interpolation and derivative operators, their discretizations  in the interior of the domain  \rf{intrnlsch}  and near boundaries \rf{bndschj}. The initial state of the model \rf{swwv-grid}  have been chosen as
\beqr
u(x,y,0)&=&\sin(\pi x/L)\sin(2\pi y/L),\; v(x,y,0)=\sin(\pi x/L)\sin(\pi y/L),
\nonumber\\
 \eta(x,y,0)&=&H_0+10\cos(\pi x/L)\cos(\pi y/(2L)).
\nonumber
\eeqr

Artificial ''observational`` data have been produced by the same model on the high resolution grid with $263\tm 263$ nodes ($h=15$ km)  with the second order approximation scheme in the interior of the domain. Assimilation experiments are performed on the $31\tm 31$ nodes grid  ($h=133$ km) with either second or fourth order approximation. 

As well as before, we use the norm of the difference $\xi$ \rf{xi} to measure the quality of assimilation. However, the solution now is not stationary  and $\xi$ varies in time also. Consequently, instead of using a value of $\xi$ at some sufficiently distant moment of time, we shall examine the evolution of  $\xi(t)$.  This evolution is compared with the evolution of the difference between the solution of the  reference model on a high resolution grid and the solution of the model on the medium resolution ($89 \tm 89$ nodes) grid. This comparison allows us to position the assimilation's result.

Several assimilation  experiments have been carried out with different number of control coefficients $K$ and $M$.  Variable assimilation window have been used in all experiments because the convergence is much more  slower in this case. Assimilating data when the solution is represented by waves, requires  more iteration with a fixed long assimilation window. Even 5000 iterations with window $T=100$ days are not sufficient to move significantly from the initial guess point and we have to use variable windows in order accelerate the convergence rate. 

Solutions of the system \rf{swwv-grid} is composed by inertia-gravity and Rossby waves. Characteristical periods of the inertia-gravity wave for the given parameters and initial conditions is of order of 17 hours   while typical period of Rossby waves is about 100 days. This difference in characteristic times makes additional advantages of using variable assimilation windows. When the window is short, of several days length,  the numerical scheme is adapted to correspond to  inertia gravity waves. And one hundred days window is more adapted to assimilate the information about Rossby waves. 

In this chapter we use the sequence $T=2,6,20,100$ days  with 50 iterations  made for each assimilation window $T$. Experiments with different $K$ and $M$ have been carried out for both second and fourth order approximation schemes in the interior of the domain.  

This evolution of the difference $\xi(t)$ is shown in \rfg{waves}. We plot this difference during 300 days time interval in order to show the evolution beyond the assimilation window and, in particular, to see whether obtained scheme is stable.     .

\begin{figure}
  \begin{center}
  \begin{minipage}[r]{1\textwidth} 
  \centerline{\includegraphics[angle=0,width=0.49\textwidth]{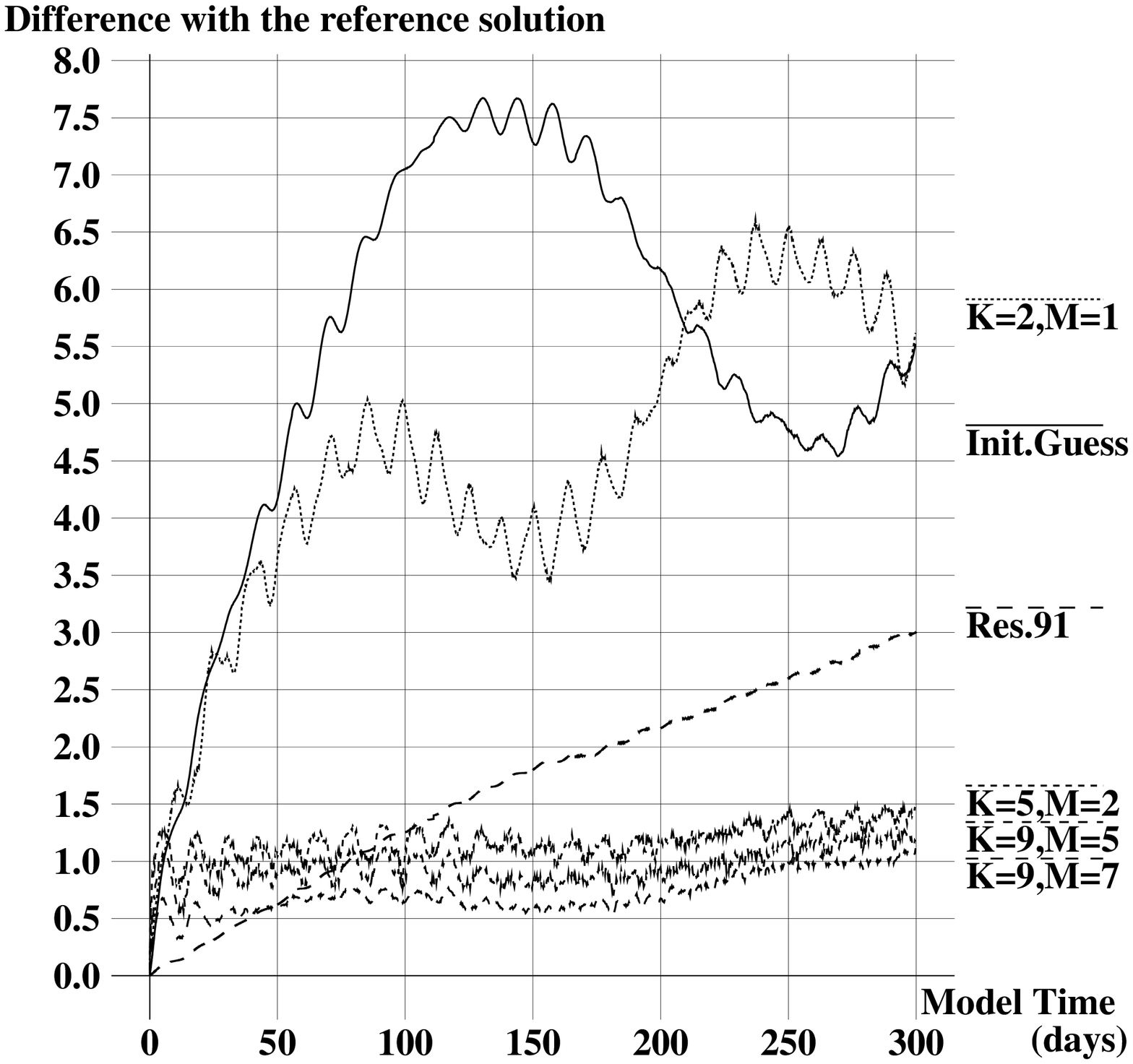}
              \includegraphics[angle=0,width=0.49\textwidth]{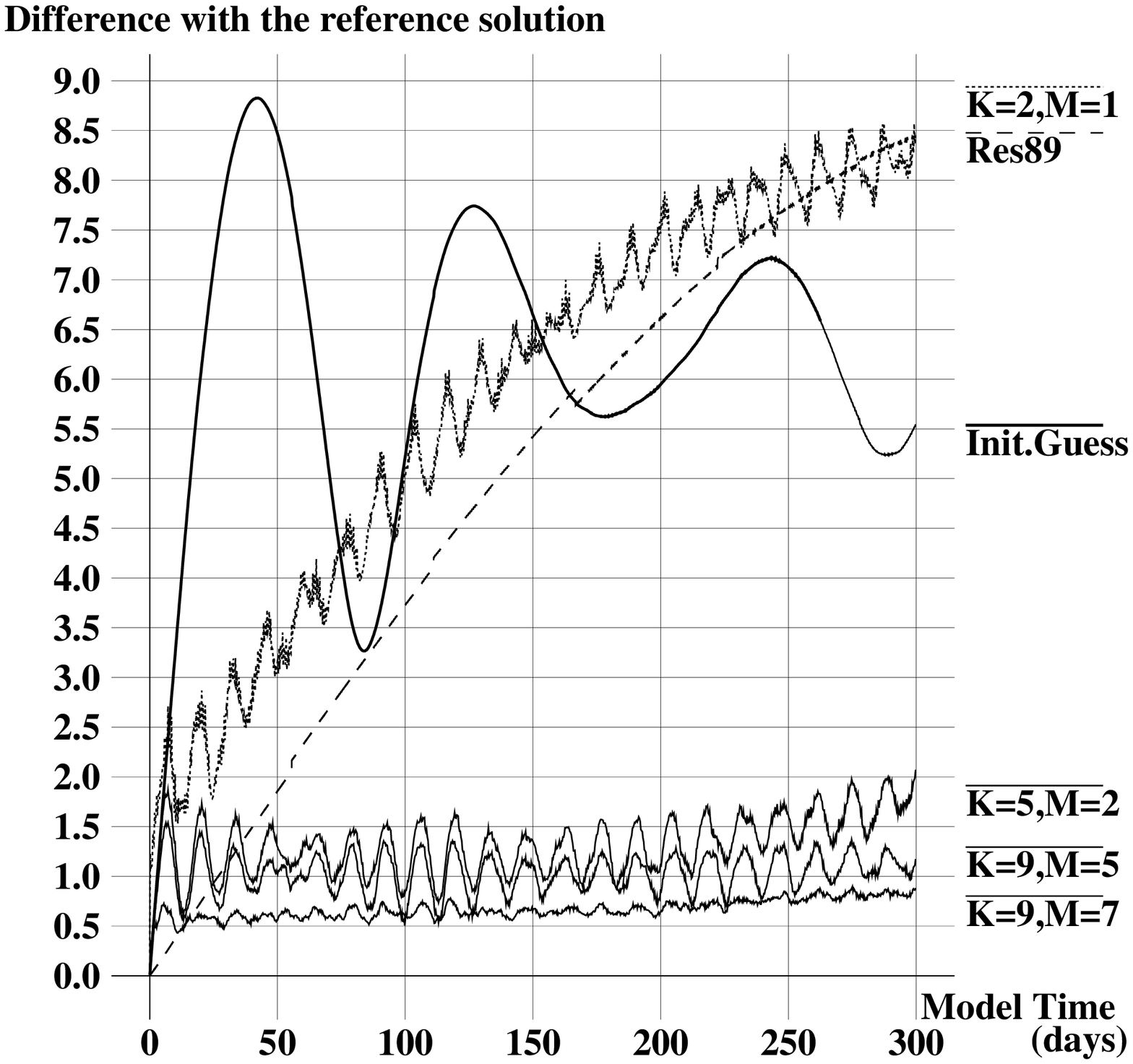}}
  \caption{Evolution of the difference $\xi(t)$ in the assimilation experiment with different $K$ and $M$ with the second (left) and fourth (right) order approximation scheme in the interior.  }
  \end{minipage} 
  \end{center} 
\refstepcounter{fig}
\label{waves}
\end{figure}

One can see that the principal error produced by the numerical scheme consists in a wrong wave speed of inertia-gravity waves. This error is visible in the \rfg{waves} because there is only one trigonometric mode in the initial conditions of the model. The phase speed error with a single mode lead to the oscillating behavior of lines representing initial guess in both left and right figures. When the second order approximation is used, the initial guess line has a minimum on the 260th day indicating that the  wave on the low-resolution grid  is exactly one period shifted with respect to the wave approximated on the high resolution grid.  Using fourth order approximation in the interior, we get this minimum earlier, on the 85th day. 

Identification of the optimal numerical scheme at just one point near boundary does not help to correct the error in wave velocity. We need to control the approximation at least at two points near boundary. However, when $M$ is equal or bigger than two, data assimilation lead to similar behavior of the solution. As it is shown in \rfg{waves} the error in velocity is corrected by optimal  numerical scheme for all $M\geq 2$, but only a little improvement is achieved by further increasing of $M$. When $M=7$, the difference with the reference solution $\xi$ is approximately two times lower  than this difference obtained with $M=2$.  Taking into account that the data assimilation is much more expensive with $M=7$, we can state that using $M=2$ is optimal in the experiments with inertia-gravity waves. We can note, that similar result has been obtained in \cite{assimbc1} in experiments with one-dimensional  wave equation. It has been shown that  if the wave composed by multiple trigonometric modes, data assimilation with $M=1$ can not correct the wave speed error because the the numerical   scheme that could be optimal is unstable.  We have to control the scheme at   two points near boundary at least. 

Comparing assimilation results with the test solution obtained using a classical scheme on the medium resolution grid ($89\tm 89$), we see the medium resolution also suffers by the error in the wave velocity. The error is, obviously, much smaller than in the initial guess case that uses a 3 times coarser grid $30\tm 30$.  Smaller error results in a slower increase of the difference $\xi$. Dashed lines in \rfg{waves}, that represent  the difference between the solution on the medium grid and on the fine grid, grow slower than solid lines, but they all reach the same maximal values. 

So far the  error in the velocity has been corrected for the solution on the coarse grid with optimal discretization near boundary, on long time scales this solution is closer to the reference solution than the solution obtained on the medium grid. 

Modifications that have been made in the discretizations of interpolation operators and derivatives near boundary are not the same as in the case of the stationary solution. Contrary to results of experiments in the previous section,   coefficients $\alpha_0$ have not moved from their initial guesses  $\alpha_0=0$ for all operators. Principal modifications have been applied to the discretizations them-self, i.e. to $\alpha_{k,m}: k\ne 0$. 

Taking into account that experiments in this section have been performed with sufficiently large numbers of control coefficients $K$ and it is difficult to understand what global effect brings each modification,  we consider the coefficients of the  Taylor expansion. 
For each linear combination like \rf{bndschj} that represent an approximation of the interpolation operator or a derivative, we calculate coefficients of the three first terms of the Taylor expansion:
\beqr
(D_x u)_{m/2}=\fr{1}{h}\sum\limits_{k=1}^{K} \alpha^{D_xu}_{k,m} u_{k-1/2} = \fr{\gamma_{1,m}}{h} u\mid_{m/2}+\gamma_{2,m}\der{u}{x}\biggl|_{m/2}+\gamma_{3,m} h\dder{u}{x}\biggl|_{m/2}
\nonumber \\
(S_x u)_{m/2}=\sum\limits_{k=1}^{K} \alpha^{S_xu}_{k,m} u_{k-1/2} = \gamma_{1,m} u\mid_{m/2}+\gamma_{2,m}h\der{u}{x}\biggl|_{m/2}+\gamma_{3,m} h^2\dder{u}{x}\biggl|_{m/2}
\eeqr
Coefficients $\gamma$ are obtained simply as $
\gamma_{l,m}=\sum\limits_{k=1}^{K} \fr{k^{l-1}}{l!} \alpha_{k,m}$
These  coefficients of the Taylor expansion calculated for $\alpha$ that have been obtained in the experiment with $K=5,M=2$ are shown in the Table \ref{tab4}. 

\tabcap{\textwidth}
\begin{table}
\caption{ Coefficients of the Taylor expansion calculated   in the experiment with $K=5,M=2$.  }
\begin{center}
\begin{small}
\begin{tabular}{|cc|ccc|ccc|}
\hline
&&\multicolumn{3}{|c|}{Second order}&\multicolumn{3}{|c|}{Fourth order}\\
Operator& m & $\gamma_1$ &$\gamma_2$ &$\gamma_3$& $\gamma_1$ &$\gamma_2$ &$\gamma_3$  \\
\hline
$(D_x u)$& 1& 0.0463  & 1.0490&  0.0393 & 0.0191 & 1.1128 & 0.2183  \\
$(D_x u)$& 2& -0.0137 & 0.8763&  0.1620 & -0.0375&  1.0366&  0.0678  \\
$(D_x p)$& 1&  0.0020 & 1.4587& -0.6313 & -0.0048&  1.2669&  0.3334 \\
$(D_x p)$& 2& 0.0041  & 1.2567&  0.0930 & -0.0065&  1.0930&  0.2289 \\
$(D_y v)$& 1& 0.1225  & 0.6699&  0.5234 & 0.0626 & 1.0514 & 0.0667 \\
$(D_y v)$& 2& -0.1252 & 1.3678& -0.3923 & -0.0145&  0.9344& -0.0462 \\
$(D_y p)$& 1& -0.0019 & 1.3599&  0.2992 &0.0013  & 0.9191 &-0.2578 \\
$(D_y p)$& 2& 0.0475  & 1.6242&  0.0138 & 0.0123 & 0.1542 &-0.4943 \\
 \hline  
$(S_x u)$& 1&1.2711   & 0.4103&  0.5558 &1.2049  & -0.3784&  0.5495 \\
$(S_x p)$& 1&0.8935   & 0.0345&  0.0006 &0.9211  & -0.2167& -0.1967 \\
$(S_x p)$& 2&0.7497   & 0.1913& -0.1252 &0.7632  &  0.1147& -0.3810 \\
$(S_y v)$& 1&1.1307   &-0.1101&  0.1195 & 1.1312 & -0.0580& -0.0097 \\
$(S_y v)$& 2&1.1259   &-0.0702&  0.2285 &1.0686  &  0.0113&  0.0139 \\
$(S_y p)$& 2&1.5921   &-1.2996&  1.2440 & 0.8134 &  0.0573&  0.0728 \\
\hline
\end{tabular}
\end{small}
\end{center}
\label{tab4}
\end{table}

 Taylor expansions show that obtained expressions  do not approximate corresponding  operators. So far, the value of $\gamma_1$ is never zero,  the derivative near boundary is always approximated with the minus first order due to the presence $\fr{\gamma_1}{h}$ in it's Taylor expansion. Values of $\gamma_1$ may be as high as 0.12, that is comparable with the coefficient in front of the first derivative $\gamma_2$, which is, in turn, may be as low as 0.15 and as high as 1.6. 

The first coefficient of the Taylor expansion of the interpolation near boundary varies also in a wide range  between 0.7 and 1.6. That means interpolation multiplies simultaneously the function by some value which is  different from 1. 

Similar phenomenon have been  seen in \cite{assimbc1} in experiments with numerically approximated  waves propagating with a wrong wave velocity.  
 The data assimilation and control of the discretization of the  derivatives near the boundary can not modify numerical wave velocity. The only way for this control to get a better solution consists in modifying  the size of the domain by deforming the grid cell adjacent to  boundary. The boundary cell  of the domain is adapted by data assimilation   to ensure the  wave with  numerical velocity propagates the modified interval in the same  time that the exact wave propagates the exact interval.  So far,  the control can not correct the error in the wave velocity, it commits another error in the interval size in order to compensate  the first one.

 \section{Conclusion}

The purpose of this paper is to study the 
variational data assimilation procedure applied for identification of the optimal parametrization of  derivatives and interpolation operators near the boundary on the example of   a linear shallow water model. 

 Comparing this procedure with now well developed data assimilation intended to identify optimal initial data, we can say there are both common points and differences as well.  
 
 Tangent \rf{tlm} and  adjoint \rf{am} models  are composed of two  types of operators:  operators acting  in the space of perturbations of $\delta u, \delta v$ and $\delta\eta$, and operators acting on the variations of the control coefficients $\delta\alpha$ directly.  The first type, that constitute the $3\tm3$ block in the matrix of \rf{tlm-mat}, governs   the evolution of a small perturbation by the model's dynamics. This term is common for any data assimilation no matter what parameter we want to identify.  The second type, \rf{R},  determines the way how the uncertainty is introduced into the model. So, if we intend to identify an optimal discretization of operators near the  boundary  for a model with an existing adjoint developed for data assimilation and identification of initial point, we can use this adjoint as $3\tm3$ block in the matrix of \rf{tlm-mat} because this part is common for any data assimilation.  However, the fourth column of this matrix \rf{R}  must be developed from the beginning because this column is specific for the particular control parameter. As we have seen, the development of the fourth column is as complex as the development of the $3\tm3$ block even for a simple linear model. For a non-linear shallow-water model, the development of the fourth column will become even more complex because non-linear terms contribute more to operators responsible for introducing of uncertainty into the model. 

Using automatic code differentiation for generation of  the adjoint model may also become more complex  in the case of  controlling the model's internal parameters. The adjoint model is not standard, the model must be  rewritten in order to satisfy the automatic differentiation requirements.

Another difference consists in the number of control parameters. The dimension of initial point of the model is usually equal to the dimension of the model's state variable. Contrary to this, when we  control boundary parametrization, the dimension of control variables is much lower than the dimension of the model's variable because  the dimension of the control  is proportional to the length of the boundary of the domain, while the dimension of the model's state relates to the area of the domain. That means the quantity of controlled parameters and the dimension of the gradient of cost function may be much lower than the quantity of variables in the models state. 

Taking into account  mentioned technical difficulties in development of the adjoint, it may be reasonable to try to calculate the gradient by some other method beginning with  the simplest finite difference method. 

As we have seen in this paper, controlling the discretization of operators in the boundary region allows to obtain better results than controlling boundary conditions of the model. The possibility to control more than two parameters, as it is the case when boundary conditions are controlled,  allows us to get lower difference with the reference model. Enlarging the boundary region, where derivatives and interpolations are  controlled,  allows us to obtain the solution on the coarse grid which is closer to the reference one than the solution on the 3 times finer grid. This is valid as for the stationary solution with under-resolved Munk layer, and for the non-stationary solution representing inertia-gravity and Rossby waves. 

In the case of  the stationary solution with  Munk layer, only one point near the boundary may be controlled allowing to avoid spurious oscillations due to under-resolved boundary layer. To obtain the solution almost indistinguishable from the reference one, we have to enlarge boundary region and to control operators at 4 nodes near boundary.

Uniform control of all  coefficients $\alpha$ all the long the boundary is not the best choice. Mass and momentum fluxes can not be balanced on the boundary and the assimilation procedure have to look for the optimum in the domain where the scheme is instable. Consequently, at least $\alpha_0$ should be allowed to be variable and particular in each point near the boundary in order to allow the  flux across the boundary to be compensated.  This requirement substantially increases the dimension of the control variables, but it is necessary to obtain a stable scheme:  we must at least allow  control coefficients  $\alpha_0$ to depend on  coordinate: on the  latitude $y$ corresponding to the index $j$ for operators $S_x,\; D_x$ and $G_x$, and on the  longitude $x$ or index $i$ for operators $S_y,\; D_y$ and $G_y$. 
 
Concerning the choice of the assimilation window, we must take into account that assimilation using short windows may provide an unstable scheme while using long windows is computationally more expensive. Moreover, better precision we want, longer windows we must use. We have seen in the experiment with the stationary solution that 
 10 iterations are  sufficient for the minimization to converge when $T=0.2$ days resulting, however,   in an unstable scheme. When  assimilation window $T=2$ days is used,  ten iterations are no longer sufficient for the minimization to converge. We need at least  50 iterations  to  bring the cost funtion close to  the minimum and to keep a stable scheme. Performing   150 supplementary iterations we win a little in the cost function but the scheme looses the stability. Assimilation window   $T=20$ days ensures the stability of the scheme during  1000 iterations, but after that  we obtain an instable scheme also. 

To combine rapid convergence of the assimilation with short $T$ and the stability of obtained scheme with long $T$ we may use  variable assimilation windows. Performing few iterations  with short $T$ we obtain the set of coefficients $\alpha$ that provides low cost function value. This set is corrected by assimilation with longer $T$ in order to ensure the stability of the scheme.

When we vary the number of control parameters $K$ and $M$, we see that parameter $K$ influences only a little  the assimilation quality. The difference between the solution produced by the model with optimal discretization of operators in the boundary region and the reference solution is very little   sensitive to the number of coefficients $\alpha$ used in the approximation of derivatives and interpolations near the boundary. On the other hand, the difference $\xi(t)$ is sensitive to $M$, that is to the number of grid nodes near boundary in which the discretization is controlled by data assimilation. In other words, to the width of the boundary region. We have seen, despite under-resolution of the Munk layer by the model's grid, optimal discretization allows us to obtain the solution as close to the reference one as we want. It is sufficient to take a sufficient $M$. 

In the experiment with waves, parameter $M$ influences also the assimilation results. One have to take $M$ at least equal to 2 in order  to compensate the error in wave velocities. 

It has been shown also that the sea surface height may be used as  the only observable variable. It is possible to reconstruct discretization of all operators using observations of $\eta$ only. However, assimilation in this case becomes 5 times more expensive due to slower convergence of the minimization procedure. 

Analyzing coefficients $\alpha$ obtained in assimilation, we can note that optimal parametrization of interpolations and derivatives near the boundary may approximate nothing in  classical sense, i.e. it may not be  valid for an arbitrary function. We have seen here that coefficients $\gamma$ of the Taylor expansion of obtained expressions may be as high as 1.6 and as low as 0.15 instead of 1. Moreover, the approximation of the derivative has minus first order.   Hence, we must take into account that  coefficients found by data assimilation do not approximate interpolations or derivatives in general. They  are valid for given model's parameters only. 

Thus, there exist a possibility to apply data assimilation technique to identification of optimal discretizations of the model's operators in the boundary region. Optimal discretization allows us to correct such errors of the  numerical scheme as under-resolved boundary layer and wrong wave velocity.

\bibliography{/global/users/kazan/text/mybibl}
 
\mkpicstoend 

\end{document}